\documentclass[prd,aps,nofootinbib,floatfix,10pt]{revtex4}
\usepackage{amsmath,graphicx,epsfig,amssymb}
\usepackage{inputenc}
\usepackage[usenames]{color}
\usepackage{ulem} %% for strike-through
\usepackage{bigstrut}
\usepackage{slashed}
\usepackage{multirow}
\usepackage{subfigure}

\allowdisplaybreaks

\begin{document}
\title{Angular Distributions for Multi-body Semileptonic Charmed Baryon Decays}
	
\author{Fei Huang$^{1}$ and
Qi-An Zhang$^{2}$~\footnote{Corresponding author: zhangqa@sjtu.edu.cn}}
\affiliation{$^{1}$INPAC, Key Laboratory for Particle Astrophysics and Cosmology (MOE), Shanghai Key Laboratory for Particle Physics and Cosmology,
		School of Physics and Astronomy, Shanghai Jiao Tong University, Shanghai
		200240, China }
\affiliation{$^{2}$Tsung-Dao Lee Institute, Shanghai Jiao Tong University, Shanghai 200240, China}

\begin{abstract}
We perform an analysis  of    angular distributions in   semileptonic decays  of charmed baryons $B_1^{(\prime)}\to B_2^{(\prime)}(\to B_3^{(\prime)}B_4^{(\prime)})\ell^+\nu_{\ell}$, where the $B_1=(\Lambda_c^+,\Xi_c^{(0,+)})$ are the SU(3)-antitriplet baryons  and $B_1'=\Omega_c^-$ is an SU(3) sextet.  We will firstly derive analytic expressions for angular distributions using   helicity amplitude technique. Based on the lattice QCD results for  $\Lambda_c^+\to\Lambda$ and $\Xi_c^0\to\Xi^-$ form factors and model calculation of the $\Omega_c^0\to\Omega^-$ transition, we predict branching fractions:  $\mathcal{B}(\Lambda_{c}^{+} \rightarrow p \pi^{-} e^{+} \nu_{e})=2.48(15)\%$, $\mathcal{B}(\Lambda_{c}^+\rightarrow p \pi^{-}\mu^{+}\nu_{\mu})=2.50(14)\%$,  $\mathcal{B}(\Xi_{c}\rightarrow \Lambda \pi^{-}e^{+}\nu_{e})=2.40(30)\%$, $\mathcal{B}(\Xi_{c}\rightarrow \Lambda \pi^{-}\mu^{+}\nu_{\nu})=2.41(30)\%$, $\mathcal{B}(\Omega_{c}\rightarrow \Lambda K^{-}e^{+}\nu_{e})=0.362(14)\%$, $\mathcal{B}(\Omega_{c}\rightarrow \Lambda K^{-}\mu^{+}\nu_{\nu})=0.350(14)\%$. Besides, we also predict  the $q^2$-dependence and angular distributions of these processes, in particular the coefficients for the $\cos n\theta_{\ell}$ ($\cos n\theta_{h}$, $\cos n\phi$) $(n=0, 1, 2, \cdots)$ terms. This work can provide a   theoretical  basis for the ongoing experiments at BESIII, LHCb and BELLE-II.
\end{abstract}
\maketitle
	
%\tableofcontents
%%%%%%%%%%%%%%%%%%%%%%%%%%%%%%%%%%%%%%%%
%%%%%%%%%%%%%%%%%%%%%%%%%%%%%%%%%%%%%%%%
\section{Introduction}

Weak decays of heavy mesons play an important role in  testing the standard model (SM), and measuring the Cabibbo-Kobayashi-Maskawa (CKM) matrix elements that describe the quark mixing and also the strength of CP violation. Moreover, any significant deviation from SM predictions for heavy meson decays will provide clues for  new physics beyond SM, and recent experimental analyses by   Belle and LHCb collaborations~\cite{Wei:2009zv,Aaij:2015esa,Aaij:2015oid,Aaij:2017vbb,Aaij:2020nrf} have revealed notable tensions between the SM predictions of such processes and data. From this viewpoint, the study of semi-leptonic decays of charmed baryons, which can  provide an ideal way to determine  the $|V_{cd}|$ and $|V_{cs}|$, and examine  the CKM unitarity $\sum_{i=d,s,b}|V_{ci}|^2=1$, is of great value.    
In the  singly-charmed baryons with $q_1q_2c$,  the two light quarks  can be decomposed as an antitriplet and a  sextet. Focusing on the ground-states with $J^P=1/2^+$, only four baryonic states,  $(\Lambda_c^+,\Xi_c^0,\Xi_c^+)$ and $\Omega_c^0$ baryons, can have measurable weak decays,   while others such like $\Xi_c'$ and $\Sigma_c'$ have strong and electromagnetic decay modes~\cite{CroninHennessy:2000bz,Ammar:2002pf,Aubert:2007bt,Yelton:2017uzv,Tanabashi:2018oca}. 	
		%	\vspace{20mm}
		
%	\item \red{Previous works.}
	
\begin{figure}[htbp]
\centering
\includegraphics[width=0.45\columnwidth]{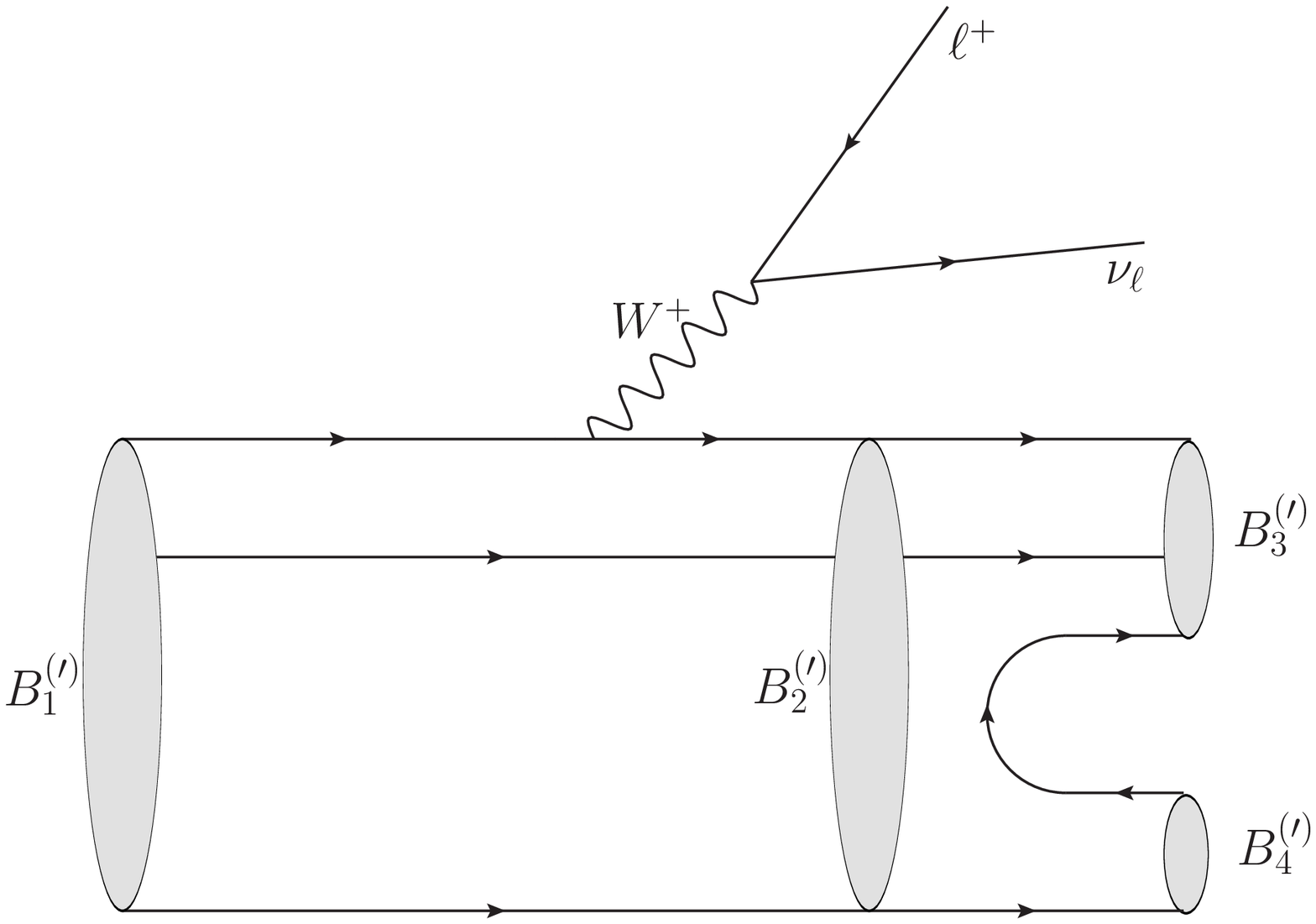}
%\caption{}
\centering
\caption{Feynman diagram of charmed baryons weak decay of $B_1^{(\prime)}\to B_2^{(\prime)}(\to B_3^{(\prime)}B_4^{(\prime)})\ell^+\nu_\ell$.}
\label{weakdecay}
\end{figure}

Among various decay modes, semileptonic decays are simplest~\cite{Richman:1995wm,Eichten:1989zv,Neubert:1993mb}, and in recent years charmed baryon decays have received  great interests from both theoretical  and experimental sides~\cite{Cheng:1991sn,Gronau:2013mza,Ablikim:2016mcr,Ablikim:2016vqd,Ablikim:2015prg,Ablikim:2017ors,Ablikim:2017iqd,Aaij:2017pgy,Aaij:2017svr,Aaij:2017nsd,Aaij:2017xva,Aaij:2017rin}.  Semileptonic decays of $\Lambda^{+}_{c}$ have been fruitfully  studied in quark model and QCD sum rules~\cite{Buras:1976dg,Gavela:1979wk,AvilaAoki:1989yi,PerezMarcial:1989yh,Hussain:1990ai,Singleton:1990ye,Efimov:1991ex,Garcia:1992qe,Cheng:1995fe,Ivanov:1996fj,Dosch:1997zx,Pervin:2005ve,Liu:2009sn,Gutsche:2015rrt,Faustov:2016yza}, and predictions for branching  fractions differ substantially. A precise measurement of branching fractions of $\Lambda_{c}$ weak decays has recently  been  reported by  BESIII collaboration: $\mathcal{B}(\Lambda_{c}\rightarrow\Lambda e^{+} \nu_{e})=0.0363$ and $\mathcal{B}(\Lambda_{c}\rightarrow\Lambda \mu^{+} \nu_{\mu})=0.0363$~\cite{Ablikim:2015prg,Ablikim:2016vqd}.
%%%
For $\Xi_{c}$, the CLEO collaboration has  measured  the ratio of branching fractions $\mathcal{B}(\Xi^{0}_{c}\rightarrow\Xi^{-} e^{+}\nu_{e})/\mathcal{B}(\Xi^{0}_{c}\rightarrow\Xi^{-} \pi^{+})$~\cite{Li:2018qak,Zyla:2020zbs}, and recently the Belle collaboration reports~\cite{Y.B.Li:2021}: $ \mathcal{B}(\Xi^{0}_{c}\rightarrow\Xi^{-} e^{+}\nu_{e})=1.72(10\pm12\pm50)\%$, $\mathcal{B}(\Xi^{0}_{c}\rightarrow\Xi^{-} \mu^{+}\nu_{\mu})=1.71(17\pm13\pm 50)\%$. 
On theoretical side, a variety of models have been developed to analyze $\Xi_{c}$ weak decays ~\cite{Zhao:2018zcb,Azizi:2011mw,Geng:2018plk,Geng:2019bfz,Faustov:2019ddj}, including a recent analysis of $\Xi_{c}\rightarrow \Xi$ transition form factors from lattice QCD~\cite{Q.A.Zhang:2021}. 
%%% 
Limited by low production rate and high background levels of current experiments,  measurements of $\Omega_{c}$  decay branching ratios are not available. In theory, branching fractions of $\Omega_{c}$ weak decays are predicted in  light-front quark model: $\mathcal{B}(\Omega^{0}_{c}\rightarrow\Omega^{-}e^{+}\nu_{e})=5.4(\pm0.2)\times 10^{-3}$~\cite{Hsiao:2020gtc}. In this work, we will make an exploration of  decay widths for semileptonic decays of charmed baryons with the LQCD results for form factors, and  in particular we for the first time derive the angular distributions for   four-body weak decays of $\Omega_{c}$.  Feynman diagrams for these decay chains are shown in Fig.~\ref{weakdecay}.
 %(1611.09696)
	
The rest of this  paper is organized as follows. In Sec.~II, we give  the theoretical framework for calculating the helicity amplitudes of charmed baryon decays, including the theoretical results of the Lorentz invariant leptonic and hadronic matrix elements. In Sec.~III, we list the differential decay widths of the three-body, as well as four-body decay formulas. Integrating out the $q^2$, we obtain   numerical results of partial decay width, as well as the illustration of momentum-transfer and angular distributions of the decay width. A brief summary will be presented in the last section.

\section{Theoretical Framework}

\subsection{Formalism}

We  will focus  on semileptonic four-body decays of both $\mathrm{SU}(3)$ $\mathbf{\bar{3}}$ and $\mathbf{6}$ ground states of singly-charmed baryon,  denoted  as $B_1^{(\prime)}\to B_2^{(\prime)}(\to B_3^{(\prime)}B_4^{(\prime)})\ell^+\nu_\ell$, where $\ell=e,\mu$ and $\nu_\ell$ are the charged and neutral leptons. For the antitriplet baryons $B_1=(\Lambda_c^+,\Xi_c^0,\Xi_c^+)$ decay, the intermediate states  $B_2$ are spin 1/2 baryon with an SU(3) octet, while the sextet $B_1^{\prime}=\Omega_c^0$  decay weakly  via the spin-3/2 decuplet intermediate baryons $B_2^{\prime} ~(\Omega^-)$.  $B_3^{(\prime)}$ and $B_4^{(\prime)}$ are the baryonic and mesonic final states, respectively. Examples of specific processes include:
\begin{align}
	\Lambda_c^+&\rightarrow\Lambda(\rightarrow p\pi^-) \ell^+\nu_\ell, \\
	\Xi_c^0&\rightarrow\Xi^-(\rightarrow \Lambda\pi^-)\ell^+\nu_\ell, \\
	\Omega_c^0&\rightarrow\Omega^-(\rightarrow \Lambda K^-)\ell^+\nu_\ell.
\end{align}

\begin{figure}[htbp]
\centering
\includegraphics[width=0.45\columnwidth]{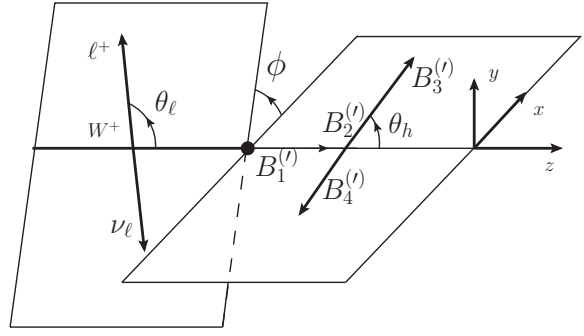}
%\caption{}
\centering
\caption{Illustration of four-body decay process $B_1^{(\prime)}\to B_2^{(\prime)}(\to B_3^{(\prime)}B_4^{(\prime)})\ell^+\nu_\ell$.}
%\label{}
\end{figure}

The effective weak Hamitonian for the semileptonic decays of charmed baryons can be written as
\begin{align}
	\mathcal{H}_{\mathrm{eff}}=\frac{G_F}{\sqrt{2}}V_{cs}\Big(\bar{s}\gamma^{\mu}(1-\gamma_5)c\Big) \Big(\bar{\nu}_{\ell}\gamma_{\mu}(1-\gamma_5)\ell\Big),
\end{align}
where $G_F$ is the Fermi constant, and $V_{cs}$ is the CKM matrix element. Based on the above effective Hamiltonian, we can obtain the decay amplitudes of $B_1^{(\prime)}\to  B_3^{(\prime)}B_4^{(\prime)}\ell^+\nu_\ell$
\begin{align}
	\mathcal{M} =&  \frac{G_F}{\sqrt{2}}V_{cs}\left\langle B_{3}^{(\prime)} B_{4}^{({\prime})}\right|\bar{s} \gamma^{\mu}\left(1-\gamma_{5}\right) c\left| B_{1}^{(\prime)}\right\rangle
	\left\langle \ell^{+} \nu_{\ell}\left|\bar{\nu}_{\ell}\gamma_{\mu}(1-\gamma_5)\ell\right| 0\right\rangle . \label{eq:fourbodyamplitude0}
\end{align}

With the  decomposition of $g_{\mu\nu}$
\begin{align}
	g_{\mu\nu}=-\sum_{\lambda}\epsilon_{\mu}^*(\lambda)\epsilon_{\nu}(\lambda)+\frac{q_{\mu}q_{\nu}}{q^2},
\end{align}
the above amplitude can be decomposed  into the Lorentz invariant hadronic and leptonic matrix elements:
 \begin{align}
	\mathcal{M}=& \frac{G_F}{\sqrt{2}}V_{cs}\left\langle B_{3}^{(\prime)} B_{4}^{({\prime})}\right|\bar{s} \gamma^{\mu}\left(1-\gamma_{5}\right) c\left| B_{1}^{(\prime)}\right\rangle
	\left\langle \ell^{+} \nu_{\ell}\left|\bar{\nu}_{\ell}\gamma^{\nu}(1-\gamma_5)\ell \right| 0\right\rangle g_{\mu\nu} \nonumber\\
	=&  \frac{G_F}{\sqrt{2}}V_{cs} \left( -\sum_{\lambda}H^{\mu}\epsilon_{\mu}^*(\lambda)\times L^{\nu}\epsilon_{\nu}(\lambda)+H^{\mu}\epsilon_{\mu}^*(t)\times L^{\nu}\epsilon_{\nu}(t) \right),
\end{align}
where the hadronic part $H^{\mu}\equiv\left\langle B_{3}^{(\prime)} B_{4}^{({\prime})}\right|\bar{s} \gamma^{\mu}\left(1-\gamma_{5}\right) c\left| B_{1}^{(\prime)}\right\rangle$ and leptonic part $L^{\nu}\equiv\left\langle \ell^{+} \nu_{\ell}\left|\bar{\nu}_{\ell}\gamma^{\nu}(1-\gamma_5)\ell\right| 0\right\rangle$, $\epsilon$ is the polarization vector of decomposed $W^+$ boson with helicity state $\lambda$ and $t$, in which  $\epsilon_{\mu}(t)\equiv q_{\mu}/\sqrt{q^2}$.

Focusing on the hadronic part, and  inserting the 1-particle completeness states of the intermediate baryon
\begin{align}
	1=\sum_{s_2}\int\frac{d^4{p}_2}{(2\pi)^4}\frac{i}{p_2^2-m_2^2+im_2\Gamma_{2}}\left|B_2^{(\prime)}(p_2,s_2)\right\rangle\left\langle B_2^{(\prime)}(p_2,s_2)\right|,
\end{align}
we find the  $H^{\mu}$ is given as: 
\begin{align}
	H^{\mu}=\sum_{s_{2}} \int \frac{d^{4} p_{2}}{(2 \pi)^{4}} \frac{i}{p_{2}^{2}-m_{2}^{2}+i m_{2} \Gamma_{2}}
	\left\langle B_{3}^{(\prime)}(p_3,s_3) B_{4}^{({\prime})}(p_4)\right.\left|B_2^{(\prime)}(p_2,s_2)\right\rangle\left\langle B_2^{(\prime)}(p_2,s_2)\right|\bar{s} \gamma^{\mu}\left(1-\gamma_{5}\right) c\left| B_{1}^{(\prime)}(p_1,s_1)\right\rangle. 
\end{align}
In the above $p_2$, $m_2$ and $\Gamma_2$ are the four-momentum, mass and decay rate of intermediate state $B_2^{(\prime)}$ respectively. 
%The initial- to intermediate-baryon transition matrix element  is denoted as $H_1^{\mu}$ while the  intermediate- to final-states hadronic matrix element is written as $H_2$. 
The $B_{1}^{(\prime)}\to B_{2}^{(\prime)}$ transition $H_1$ can be parameterized as hadronic form factors, while the $B_{2}^{(\prime)}\to B_{3}^{(\prime)}B_{4}^{(\prime)}$ transition, with one final-state $B_{4}^{(\prime)}$ is pseudoscalar meson, can be reduced as
\begin{align}
&	\left\langle B_{3}^{(\prime)}(p_3,s_3) B_{4}^{({\prime})}(p_4) \right.\left| B_2^{(\prime)}(p_2,s_2)\right\rangle_{H} 
= \left\langle B_{3}^{(\prime)}(p_3,s_3) B_{4}^{({\prime})}(p_4)\left|-i \int d^{4} x \mathcal{H}_{\mathrm{int}}(x)\right| B_2^{(\prime)}(p_2,s_2)\right\rangle_{I} \nonumber\\
&= \left\langle B_{3}^{(\prime)}(p_3,s_3) B_{4}^{({\prime})}(p_4)\left|-i \int d^{4} x 
	\Big( ig_{H_2}\bar{\psi}_{B_{3}}(x) \phi_{B_{4}}(x)\left(A-B \gamma_{5}\right) \psi_{B_{2}}(x) \Big)\right| B_2^{(\prime)}(p_2,s_2)\right\rangle \nonumber\\
&= (2 \pi)^{4} \delta^{4}\left(p_{3}+p_{4}-p_{2}\right) g_{H_2}\bar{u}_{B_{3}}\left(p_{3}, s_{3}\right)\left(A-B \gamma_{5}\right) u_{B_2}\left(p_{2}, s_{2}\right),
\end{align}
where the subscript ``$H$'' and ``$I$'' denote Heisenberg and interaction representation matrix element respectively, $g_{H_2}$ represents the coupling constant of hadronic vertex $B_{2}^{(\prime)}\to B_{3}^{(\prime)}B_{4}^{(\prime)}$ , and factors $A$ and $B$ are constants weighting the contributions from scalar and pesudoscalar density operators~\cite{Zyla:2020zbs}.

Therefore, the decay amplitude in Eq.(\ref{eq:fourbodyamplitude0}) can be expressed as a convolution of the Lorentz invariant leptonic part $L(s_{\ell}, s_{\nu}, s_W)$ and two hadronic parts $H_1(s_1,s_2,s_W)$, $H_2(s_2,s_3)$:
\begin{align}
	i\mathcal{M}\left(B_1^{(\prime)}\to B_3^{(\prime)}B_4^{(\prime)}\ell^+\nu_\ell\right)=&\sum_{s_2} i\mathcal{M}\left(B_1^{(\prime)}\to B_2^{(\prime)}\ell^+\nu_\ell\right)\times i\mathcal{M}\left(B_2^{(\prime)}\to B_3^{(\prime)}B_4^{(\prime)}\right) \times \frac{i}{p_{2}^{2}-m_{2}^{2}+i m_{2} \Gamma_{2}} \nonumber\\
	=&\sum_{s_2} \left( i\frac{G_F}{\sqrt{2}}V_{cs}\sum_{s_W=\{\pm,0,t\}} H_1(s_1,s_2,s_W) \times L(s_{\ell}, s_{\nu},s_W)  \right) \times H_2(s_2,s_3) \nonumber\\
	&\qquad\times \frac{i}{p_{2}^{2}-m_{2}^{2}+i m_{2} \Gamma_{2}} \nonumber\\
	=& -\frac{G_FV_{cs}}{\sqrt{2}}\frac{1}{p_{2}^{2}-m_{2}^{2}+i m_{2} \Gamma_{2}} \sum_{s_2}\sum_{s_W=\{\pm,0,t\}}H_1(s_1,s_2,s_W)\times L(s_{\ell}, s_{\nu},s_W) \times H_2(s_2,s_3), \label{eq:totaldecayamplitude}
\end{align}
with
\begin{align}
	L(s_{\ell}, s_{\nu},s_W)\equiv& ~\bar{u}_{\nu}(p_{\nu},s_{\nu}) \gamma^{\nu}\left(1-\gamma_{5}\right) v_{\ell}(p_{\ell},s_{\ell}) \varepsilon_{\nu}(s_W), \label{eq:leptonicpartori}\\
	H_1(s_1,s_2,s_W)\equiv &\left\langle B_2^{(\prime)}(p_2,s_2)\right|\bar{s} \gamma^{\mu}\left(1-\gamma_{5}\right) c\left| B_{1}^{(\prime)}(p_1,s_1)\right\rangle \epsilon_{\mu}^*(s_W), \label{eq:hadronicpartH1ori} \\
	H_2(s_2,s_3)\equiv & ~ g_{H_2}\bar{u}_{B_{3}}\left(p_{3}, s_{3}\right)\left(A-B \gamma_{5}\right) u_{B_2}\left(p_{2}, s_{2}\right).\label{eq:hadronicpartH2ori}
\end{align}

Averaging the spin of initial-state and summing over the spins of final-states, we obtain  the squared amplitude as
\begin{align}
	\frac{1}{2}\overline{\left|i\mathcal{M}\left(B_1^{(\prime)}\to B_3^{(\prime)}B_4^{(\prime)}\ell^+\nu_\ell\right)\right|^2}=&\frac{G_F^2|V_{cs}|^2}{2}\frac{1}{(q_2^2-m_2^2)^2+m_2^2\Gamma_2^2}\frac{1}{2}\sum_{s_1}\sum_{s_3,s_l,s_{\nu}} \nonumber\\
	&\times\quad\left|  \sum_{s_2}\sum_{s_W=\{\pm,0,t\}}H_1(s_1,s_2,s_W)\times L(s_{\ell}, s_{\nu},s_W) \times H_2(s_2,s_3)\right|^2. \label{eq:fourbodyamplitude}
\end{align}

\subsection{Kinematics}

With the abbreviations
\begin{align}
	s_{\pm}=(m_1\pm m_2)^2-q^2,~\lambda(m_1,m_2,q)=s_+s_-,
\end{align} 
the momentum of initial- and final-states in the subprocesses can be set as follows. 
\begin{enumerate}
	%%%%%% B_1-> B_2W
	\item $B_1^{(\prime)}\rightarrow B_2^{(\prime)}W^+$ subprocess: in the rest frame of initial-state $B_1^{(\prime)}$, suppose $B_2^{(\prime)}$ moves along the positive $z$-direction  ($\theta=0,\phi=0$)  and $W^+$ boson along the negative $z$-direction ($\theta=\pi,\phi=\pi$), so we have
	\begin{align}
	p_1^{\mu}=(m_1,0,0,0),~p_2^{\mu}=(E_2,0,0,|\vec{p}_2|),~q^{\mu}=(E_W,0,0,-|\vec{p}_2|), \label{eq:hadronH1momentum}
\end{align}
with
\begin{align}
	E_1=\frac{m_1^2+m_2^2-q^2}{2m_1},~E_W=\frac{m_1^2-m_2^2+q^2}{2m_1},~ |\vec{p}_2|=\frac{\sqrt{\lambda(m_1,m_2,q)}}{2m_1}.
\end{align}
%%%%%%% B_2 -> B_3B_4
	\item $B_2^{(\prime)}\rightarrow B_3^{(\prime)}B_4^{(\prime)}$ subprocess: in the rest frame of $B_2^{(\prime)}$, suppose the final-state baryon $B_3^{(\prime)}$ moves along  ($\theta=\theta_h,\phi=0$) direction, thereby the meson $B_4^{(\prime)}$ moves along ($\theta=\pi-\theta_h,\phi=\pi$) direction, we have
\begin{align}
	p_2^{\mu}=(m_2,0,0,0), ~ p_3^{\mu}=(E_3, |\vec{p}_3|\sin\theta_h, 0,|\vec{p}_3|\cos\theta_h),~ p_4^{\mu}=(E_4, -|\vec{p}_3|\sin\theta_h, 0,-|\vec{p}_3|\cos\theta_h),
\end{align}
with 
\begin{align}
	E_3=\frac{m_2^2+m_3^2-m_4^2}{2m_2},~E_4=\frac{m_2^2-m_3^2+m_4^2}{2m_2},~ |\vec{p}_3|=\frac{\sqrt{\lambda(m_2,m_3,m_4)}}{2m_2}.
\end{align}
%%%%%%%%% W -> lnu
\item $W^+\rightarrow \ell^+\nu_l$ subprocess: in the rest frame of $W^+$, suppose the charged lepton moves along ($\theta=\theta_l,\phi=\phi$) direction, thereby the neutrino along ($\theta=\pi-\theta_l,\phi=\pi+\phi$) direction, we have
\begin{align}
	q^{\mu}=&\left(\sqrt{q^2},0,0,0\right), ~p_l^{\mu}=(E_l, |\vec{p}_l|\sin\theta_l\cos\phi,  |\vec{p}_l|\sin\theta_l\sin\phi,  |\vec{p}_l|\cos\theta_l), \nonumber\\
	 &~~~~p_{\nu}^{\mu}=(E_\nu, -|\vec{p}_l|\sin\theta_l\cos\phi,  -|\vec{p}_l|\sin\theta_l\sin\phi,  -|\vec{p}_l|\cos\theta_l), \label{eq:leptonsmomentum}
\end{align}
with
\begin{align}
	E_l=\frac{q^2+m_l^2}{2\sqrt{q^2}},~E_\nu=\frac{q^2-m_l^2}{2\sqrt{q^2}}, ~|\vec{p}_l|=\frac{q^2-m_l^2}{2\sqrt{q^2}}.
\end{align}
	\end{enumerate}

For the phase space of $n$-body decays, the four-body one can be generated by the two-body sub-processes recursively. The two-body phase space can be expressed as
\begin{align}
	d\Phi_2(p\to p_1p_2)=\frac{1}{(2\pi)^5}\frac{|\vec{p}_1|d\cos\theta}{4\sqrt{\hat{s}}}, \label{eq:2bodyphasespace}
\end{align}
where $\sqrt{\hat{s}}=\sqrt{p^2}$ is the center-of-mass energy, $\theta$ is the angle between two final-states. Based on this, the three-body phase space of $B_1^{(\prime)}\rightarrow B_2^{(\prime)}\ell^+\nu_{\ell}$ can be written as 
\begin{align}
    d\Phi_3\left(B_1^{(\prime)}\rightarrow B_2^{(\prime)} \ell^+\nu_{\ell}\right)=&(2\pi)^3dq^2\times d\Phi_2(W^+\rightarrow \ell^+\nu_{\ell})\times d\Phi_2(B_1^{(\prime)}\rightarrow B_2^{(\prime)}W^+) \nonumber\\
	=&\frac{(1-\hat{m}_l^2)\sqrt{\lambda(m_1,m_2,q)}}{(2\pi)^{7}32m_1^2}dq^2d\cos\theta_l, \label{eq:threebodyphasespace}
 \end{align}
 where $\hat{m}_l=m_l/\sqrt{q^2}$. Similarly, the total four-body phase space is
  \begin{align}
	d\Phi_4\left(B_1^{(\prime)}\to B_3^{(\prime)}B_4^{(\prime)}\ell^+\nu_\ell\right)=&(2\pi)^3dp_2^2\times  d\Phi_3\left(B_1^{(\prime)}\rightarrow B_2^{(\prime)} \ell^+\nu_{\ell}\right)\times d\Phi_2\left(B_2^{(\prime)}\rightarrow B_3^{(\prime)}B_4^{(\prime)}\right) \nonumber\\
	=&\frac{(1-\hat{m}_l^2)\sqrt{\lambda(m_1,m_2,q)\lambda(m_2,m_3,m_4)}}{(2\pi)^{10}256m_1^2m_2^2}dq^2dp_2^2d\cos\theta_hd\cos\theta_ld\phi. \label{eq:fourbodyphasespace}
 \end{align}
 
 Note that the integration variable $p_2^2$ is artificially introduced from the insertion of intermediate state $ B_2^{(\prime)}$, and in the narrow-width limit, this integration will be conducted as
 \begin{align}
 	\int dq_2^2\frac{m_2\Gamma_2}{\pi}\frac{1}{(q_2^2-m_2^2)^2+m_2^2\Gamma_2^2}=1,
 \end{align}
 while 
 \begin{align}
 	\int dq_2^2\frac{m_2\Gamma\left(B_2^{(\prime)}\rightarrow B_3^{(\prime)}B_4^{(\prime)}\right)}{\pi}\frac{1}{(q_2^2-m_2^2)^2+m_2^2\Gamma_2^2}=\mathcal{B}\left(B_2^{(\prime)}\rightarrow B_3^{(\prime)}B_4^{(\prime)}\right),
 \end{align}
where $\Gamma\left(B_2^{(\prime)}\rightarrow B_3^{(\prime)}B_4^{(\prime)}\right)$ and $\mathcal{B}\left(B_2^{(\prime)}\rightarrow B_3^{(\prime)}B_4^{(\prime)}\right)$ are total width and branching fraction of the subprocess $B_2^{(\prime)}\rightarrow B_3^{(\prime)}B_4^{(\prime)}$ respectively. 

Combining the four-body phase space in Eq.(\ref{eq:fourbodyphasespace}) and squared amplitude in Eq.(\ref{eq:fourbodyamplitude}), we can write down the differential decay width of four-body process $B_1^{(\prime)}\rightarrow B_3^{(\prime)}B_4^{(\prime)}\ell^+\nu_{\ell}$:
\begin{align}
	\frac{d\Gamma(B_1^{(\prime)}\rightarrow B_3^{(\prime)}B_4^{(\prime)}\ell^+\nu_{\ell})}{dq^2d\cos\theta_hd\cos\theta_ld\phi}
=&\frac{G_F^2|V_{cs}|^2(1-\hat{m}_l^2)\sqrt{\lambda(m_1,m_2,q)\lambda(m_2,m_3,m_4)}}{(2\pi)^{5}4096m_1^3m_2^3} \frac{\mathcal{B}\left(B_2^{(\prime)}\rightarrow B_3^{(\prime)}B_4^{(\prime)}\right)}{\Gamma\left(B_2^{(\prime)}\rightarrow B_3^{(\prime)}B_4^{(\prime)}\right)} \nonumber\\
	&\times\sum_{s_1}\sum_{s_3,s_l,s_{\nu}} \left|  \sum_{s_2}\sum_{s_W=\{\pm,0,t\}}H_1(s_1,s_2,s_W)\times L(s_{\ell}, s_{\nu},s_W) \times H_2(s_2,s_3)\right|^2. \label{eq:general4bodydecaywidth}
\end{align}

%%%%%%%%%%%%%%%%%%%%%%%%%%%%%%%%%%%%%%%%%%%%%%%%%%%%%%%%%%%%%%%

\subsection{Leptonic part} \label{sec:leptonpart}

In this section, we will focus on the Lorentz invariant leptonic part defined in Eq.(\ref{eq:leptonicpartori}):
\begin{align}
	L(s_{\ell}, s_{\nu},s_W)\equiv& ~\bar{u}_{\nu}(p_{\nu},s_{\nu}) \gamma^{\nu}\left(1-\gamma_{5}\right) v_{\ell}(p_{\ell},s_{\ell}) \varepsilon_{\nu}(s_W).
\end{align}

Combining each components of spinors of leptons and polarization vectors of $W^+$ boson, we can obtain the following non-vanishing matrix elements:
\begin{align}
L\left(s_{\ell}=+\frac{1}{2}, s_{\nu}=-\frac{1}{2}, s_{W}=0\right) &=-\sqrt{2} \sin \theta_{\ell} N_{\ell}, \label{eq:leptonicME1}\\
L\left(s_{\ell}=+\frac{1}{2}, s_{\nu}=-\frac{1}{2}, s_{W}=+1\right) &=-e^{-i \phi}\left(1-\cos \theta_{\ell}\right) N_{\ell}, \\
L\left(s_{\ell}=+\frac{1}{2}, s_{\nu}=-\frac{1}{2}, s_{W}=-1\right) &=-e^{i \phi}\left(1+\cos \theta_{\ell}\right) N_{\ell}, \\
L\left(s_{\ell}=-\frac{1}{2}, s_{\nu}=-\frac{1}{2}, s_{W}=0\right) &=-\sqrt{2} \hat{m}_{\ell} \cos \theta_{\ell} N_{\ell} \\
L\left(s_{\ell}=-\frac{1}{2}, s_{\nu}=-\frac{1}{2}, s_{W}=+1\right) &=-\hat{m}_{\ell} e^{-i \phi} \sin \theta_{\ell} N_{\ell}, \\
L\left(s_{\ell}=-\frac{1}{2}, s_{\nu}=-\frac{1}{2}, s_{W}=-1\right) &=\hat{m}_{\ell} e^{i \phi} \sin \theta_{\ell} N_{\ell}, \\
L\left(s_{\ell}=-\frac{1}{2}, s_{\nu}=-\frac{1}{2}, s_{W}=t\right) &=\sqrt{2} \hat{m}_{\ell} N_{\ell}, \label{eq:leptonicME2}
\end{align}
where the factor $N_{\ell}=i\sqrt{2\left(q^2-m_l^2\right)}$.

%%%%%%%%%%%%%%%%%%%%%%%%%%%%%%%%%%%%%%%%%%%%%%%%%%%%%%%%%%%%%%%

\subsection{Hadronic part $H_1$ with spin-1/2 intermediate state} \label{sec:hadronH1with12}

If $B_1$ is a singly-charmed baryon antitriplet, the transition matrix elements	with weak current $J^{\mu}=V^{\mu}-A^{\mu}=\bar{s}\gamma^{\mu}(1-\gamma_5)c$ in Eq.(\ref{eq:hadronicpartH1ori}) can be parameterized as~\cite{Mott:2011cx,Pervin:2005ve} 
\begin{align}
\left\langle B_{2}\left(p_{2}, s_{2}\right)\left|V^{\mu}\right| B_{1}\left(p_{1}, s_{1}\right)\right\rangle=&\bar{u}\left(p_{2}, s_{2}\right)\left[\gamma^{\mu} f_{1}\left(q^{2}\right)+i \sigma^{\mu \nu} \frac{q_{\nu}}{m_{1}} f_{2}\left(q^{2}\right)+\frac{q^{\mu}}{m_{1}} f_{3}\left(q^{2}\right)\right] u\left(p_{1}, s_{1}\right), \label{spinhalfV}\\
\left\langle B_{2}\left(p_{2}, s_{2}\right)\left|A^{\mu}\right| B_{1}\left(p_{1}, s_{1}\right)\right\rangle=&\bar{u}\left(p_{2}, s_{2}\right)\left[\gamma^{\mu} g_{1}\left(q^{2}\right)+i \sigma^{\mu \nu} \frac{q_{\nu}}{m_{1}} g_{2}\left(q^{2}\right)+\frac{q^{\mu}}{m_{1}} g_{3}\left(q^{2}\right)\right] \gamma_{5} u\left(p_{1}, s_{1}\right),\label{spinhalfA}
\end{align}
with the momentum transfer $q^{\mu}=p_{1}^{\mu}-p_{2}^{\mu}$ and $\sigma^{\mu\nu}=i[\gamma^{\mu},\gamma^{\nu}]/2$. The form factors $f_i$ and $g_{i}$ are functions of $q^2$, and the relations between these form factors and other parametrizations  are collected in appendix.

By combining each spin components of $B_1$, $B_2$ and $W^+$, we give  the non-zero terms of $H_1(s_1,s_2,s_W)$ as
 \begin{align}
H_{1 V}\left(s_1=-\frac{1}{2}, s_2=\frac{1}{2}, s_W=1\right) &=H_{1 V}\left(s_1=\frac{1}{2}, s_2=-\frac{1}{2}, s_W=-1\right) \nonumber\\  \label{eq:hadronicH1ME1}
&=\sqrt{2 s_{-}}\left[f_{1}(q^2)-\frac{m_{1}+m_{2}}{m_{1}} f_{2}(q^2)\right], \\
H_{1 V}\left(s_1=\frac{1}{2}, s_2=\frac{1}{2}, s_W=0\right) &=H_{1 V}\left(s_1=-\frac{1}{2}, s_2=-\frac{1}{2}, s_W=0\right) \nonumber\\
&=\sqrt{\frac{s_{-}}{q^{2}}}\left[\left(m_{1}+m_{2}\right) f_{1}(q^2)-\frac{q^{2}}{m_{1}} f_{2}(q^2)\right], \\
H_{1 V}\left(s_1=\frac{1}{2}, s_2=\frac{1}{2}, s_W=t\right) &=H_{1 V}\left(s_1=-\frac{1}{2}, s_2=-\frac{1}{2}, s_W=t\right) \nonumber\\
&=\sqrt{\frac{s_{+}}{q^{2}}}\left[\left(m_{1}-m_{2}\right) f_{1}(q^2)+\frac{q^{2}}{m_{1}} f_{3}(q^2)\right], \\
H_{1 A}\left(s_1=-\frac{1}{2}, s_2=\frac{1}{2}, s_W=1\right) &=-H_{1 A}\left(s_1=\frac{1}{2}, s_2=-\frac{1}{2}, s_W=-1\right) \nonumber\\
&=\sqrt{2 s_{+}}\left[g_{1}(q^2)+\frac{m_{1}-m_{2}}{m_{1}} g_{2}(q^2)\right], \\
H_{1 A}\left(s_1=\frac{1}{2}, s_2=\frac{1}{2}, s_W=0\right) &=-H_{1 A}\left(s_1=-\frac{1}{2}, s_2=-\frac{1}{2}, s_W=0\right) \nonumber\\
&=\sqrt{\frac{s_{+}}{q^{2}}}\left[\left(m_{1}-m_{2}\right) g_{1}(q^2)+\frac{q^{2}}{m_{1}} g_{2}(q^2)\right], \\
H_{1 A}\left(s_1=\frac{1}{2}, s_2=\frac{1}{2}, s_W=t\right) &=-H_{1 A}\left(s_1=-\frac{1}{2}, s_2=-\frac{1}{2}, s_W=t\right) \nonumber\\
&=\sqrt{\frac{s_{-}}{q^{2}}}\left[\left(m_{1}+m_{2}\right) g_{1}(q^2)-\frac{q^{2}}{m_{1}} g_{3}(q^2)\right]. 
\end{align}
Reversing the helicities give the same results for the vector current, but an opposite sign exists for the axial-vector current. The total amplitudes are
 \begin{align}
  H_{1}(s_1,s_2,s_W)=  H_{1V}(s_1,s_2,s_W)-  H_{1A}(s_1,s_2,s_W)\; .  \label{eq:hadronicH1ME2}
 \end{align}

%%%%%%%%%%%%%%%%%%%%%%%%%%%%%%%%%%%%%%%%%%%%%%%%%%%%%%%%%%%%%%%
\subsection{Hadronic part $H_1$ with spin-3/2 intermediate state} \label{sec:hadronH1with32}

The charmed baryons $B_1'=\Omega_c^0$ will weakly decay into spin-3/2 baryon decuplets $\Omega^-$, and the transition matrix elements for this process can be parametrized as
 \begin{align}
\left\langle B_{2}^{\prime}\left(p_{2}, s_{2}\right)\left|V^{\mu}\right| B_{1}^{\prime}\left(p_{1}, s_{1}\right)\right\rangle=\bar{U}_{\alpha}\left(p_{2}, s_2\right)[& \gamma^{\mu} p_{1}^{\alpha} \frac{f_{1}\left(q^{2}\right)}{m_{1}}+\frac{f_{2}\left(q^{2}\right)}{m_{1}^{2}} p_{1}^{\alpha} p_{1}^{\mu} \label{spinonehalfV}\nonumber\\
&\left.+\frac{f_{3}\left(q^{2}\right)}{m_{1} m_{2}} p_{1}^{\alpha} p_{2}^{\mu}+f_{4}\left(q^{2}\right) g^{\alpha \mu}\right] \gamma^{5} u\left(p_{1}, s_{1}\right), \\
\left\langle B_{2}^{\prime}\left(p_{2}, s_2\right)\left|A^{\mu}\right| B_{1}^{\prime}\left(p_{1}, s_{1}\right)\right\rangle=\bar{U}_{\alpha}\left(p_{2}, s_2\right)[& \gamma^{\mu} p_{1}^{\alpha} \frac{g_{1}\left(q^{2}\right)}{m_{1}}+\frac{g_{2}\left(q^{2}\right)}{m_{1}^{2}} p_{1}^{\alpha} p_{1}^{\mu} \nonumber\\
&\left.+\frac{g_{3}\left(q^{2}\right)}{s_2m_{1} m_{2}} p_{1}^{\alpha} p_{2}^{\mu}+g_{4}\left(q^{2}\right) g^{\alpha \mu}\right] u\left(p_{1}, s_{1}\right). \label{spinonehalfA}
\end{align}
The definition of vectorial spinor $U_{\alpha}(\vec{p},S_z)$ for spin-$3/2$ baryon is shown in appendix. Therefore, the non-zero terms of $H_1(s_1,s_2,s_W)$ are collected as 
 \begin{align}
 %%%
H_{1V}\left(s_{1}=\frac{1}{2}, s_2=\frac{3}{2}, s_{W}=1\right) &=-H_{1V}\left(s_{1}=-\frac{1}{2}, s_2=-\frac{3}{2}, s_{W}=-1\right) \nonumber\\ \label{eq:hadronic32H1start}
&=\sqrt{s_{-}} f_{4}\left(q^{2}\right), \\
%%%
H_{1V}\left(s_{1}=-\frac{1}{2}, s_2=\frac{1}{2}, s_{W}=1\right) &=-H_{1V}\left(s_{1}=\frac{1}{2}, s_2=-\frac{1}{2}, s_{W}=-1\right) \nonumber\\
&=\sqrt{\frac{s_{-}}{3}}\left(\frac{s_{+}}{m_{1} m_{2}} f_{1}\left(q^{2}\right)-f_{4}\left(q^{2}\right)\right), \\
%%%
H_{1V}\left(s_{1}=\frac{1}{2}, s_2=\frac{1}{2}, s_{W}=0\right)&=-H_{1V}\left(s_{1}=-\frac{1}{2}, s_2=-\frac{1}{2}, s_{W}=0\right) \nonumber\\
&= \sqrt{\frac{s_{-}}{6 q^{2}}}\left[\frac{\left(m_{1}-m_{2}\right) s_{+}}{m_{1} m_{2}} f_{1}\left(q^{2}\right)\right. \nonumber\\
&- \frac{\lambda\left(m_{1}, m_{2}, q\right)}{2 m_{1} m_{2}}\left(\frac{1}{m_{1}} f_{2}\left(q^{2}\right)+\frac{1}{m_{2}} f_{3}\left(q^{2}\right)\right) \nonumber\\
&\left.-\frac{m_{1}^{2}-m_{2}^{2}-q^{2}}{m_{2}} f_{4}\left(q^{2}\right)\right], \\
%%%
H_{1V}\left(s_{1}=\frac{1}{2}, s_2=\frac{1}{2}, s_{W}=t\right)&=-H_{1V}\left(s_{1}=-\frac{1}{2}, s_2=-\frac{1}{2}, s_{W}=t\right) \nonumber\\
&=\sqrt{\frac{s_{+}}{q^{2}}} \frac{s_{-}}{\sqrt{6} m_{2}}\left[\frac{m_{1}+m_{2}}{m_{1}} f_{1}\left(q^{2}\right)\right. \nonumber\\
&-\frac{m_{1}^{2}-m_{2}^{2}+q^{2}}{2 m_{1}^{2}} f_{2}\left(q^{2}\right) \nonumber\\
&\left.-\frac{m_{1}^{2}-m_{2}^{2}-q^{2}}{2 m_{1} m_{2}} f_{3}\left(q^{2}\right)-f_{4}\left(q^{2}\right)\right], \\
%%%
H_{1A}\left(s_{1}=\frac{1}{2}, s_2=\frac{3}{2}, s_{W}=1\right) &=H_{1A}\left(s_{1}=-\frac{1}{2}, s_2=-\frac{3}{2}, s_{W}=-1\right) \nonumber\\
&=-\sqrt{s_{+}} g_{4}\left(q^{2}\right), \\
%%%
H_{1A}\left(s_{1}=-\frac{1}{2}, s_2=\frac{1}{2}, s_{W}=1\right) &=H_{1A}\left(s_{1}=\frac{1}{2}, s_2=-\frac{1}{2}, s_{W}=-1\right) \nonumber\\
&=\sqrt{\frac{s_{+}}{3}}\left(\frac{s_{-}}{m_{1} m_{2}} g_{1}\left(q^{2}\right)-g_{4}\left(q^{2}\right)\right), \\
%%%
H_{1A}\left(s_{1}=\frac{1}{2}, s_2=\frac{1}{2}, s_{W}=0\right) &=H_{1A}\left(s_{1}=-\frac{1}{2}, s_2=-\frac{1}{2}, s_{W}=0\right) \nonumber\\
&=\sqrt{\frac{s_{+}}{6 q^{2}}}\left[\frac{\left(m_{1}+m_{2}\right) s_{-}}{m_{1} m_{2}} g_{1}\left(q^{2}\right)\right. \nonumber\\
&+\frac{\lambda\left(m_{1}, m_{2}, q\right)}{2 m_{1} m_{2}}\left(\frac{1}{m_{1}} g_{2}\left(q^{2}\right)+\frac{1}{m_{2}} g_{3}\left(q^{2}\right)\right) \nonumber\\
&\left.+\frac{m_{1}^{2}-m_{2}^{2}-q^{2}}{m_{2}} g_{4}\left(q^{2}\right)\right], \\
%%%
H_{1A}\left(s_{1}=\frac{1}{2}, s_2=\frac{1}{2}, s_{W}=t\right)&=H_{1A}\left(s_{1}=-\frac{1}{2}, s_2=-\frac{1}{2}, s_{W}=t \right) \nonumber\\
&=\sqrt{\frac{s_{-}}{q^{2}}}  \frac{s_{+}}{\sqrt{6} m_{2}}\left[\frac{m_{1}-m_{2}}{m_{1}} g_{1}\left(q^{2}\right)\right. \nonumber\\
&+\frac{m_{1}^{2}-m_{2}^{2}+q^{2}}{2 m_{1}^{2}} g_{2}\left(q^{2}\right) \nonumber\\
&+\left.\frac{m_{1}^{2}-m_{2}^{2}-q^{2}}{2 m_{1} m_{2}} g_{3}\left(q^{2}\right)+g_{4}\left(q^{2}\right)\right],\label{eq:hadronic32H1end}
\end{align}
and the total transition amplitudes give
\begin{align}
	H_1\left(s_{1}, s_2, s_{W}\right)=H_{1A}\left(s_{1}, s_2, s_{W}\right)-H_{1V}\left(s_{1}, s_2, s_{W}\right).
\end{align}

%%%%%%%%%%%%%%%%%%%%%%%%%%%%%%%%%%%%

\subsection{Hadronic part $H_2$ with spin-1/2 intermediate state} \label{sec:hadronH2with12}
	
The amplitude for a spin-1/2 charm baryon $B_2$ decaying into a spin-1/2 baryon $B_3$ and a spin-0 meson $B_4$ can be written in the form
\begin{align}
	i\mathcal{M}=iG_Fm_4^2\bar{u}_{B_3}(p_3,s_3)(A-B\gamma_5)u_{B_2}(p_2,s_2),
\end{align}
which corresponds to the hadronic coupling constant $g_H=G_Fm_4^2$ in Eq.(\ref{eq:hadronicpartH2ori}). $A$ and $B$ are factors weighting the contributions from scalar and pesudoscalar operators~\cite{Commins:book}. It is convenient to introduce the asymmetry parameter
\begin{align}
	\alpha=\frac{2 \operatorname{Re}\left[s^{*} r\right]}{|s|^{2}+|r|^{2}},
\end{align}
with $s=A$ and $r=B\times|\vec{p}_3|/(E_3+m_3)$, and follow $(s\pm r)^2=(|s|^{2}+|r|^{2})(1\pm\alpha)$.

Therefore, combining each spin components of $B_2$ and $B_3$, we get the following hadronic matrix elements $H_2(s_2,s_3)$:
\begin{align}
H_{2}\left(s_{2}=\frac{1}{2}, s_{3}=\frac{1}{2}\right) &=N_{h}(s+r) \cos \left(\theta_{h} / 2\right), \label{eq:hadronic12H2ME1}\\
H_{2}\left(s_{2}=\frac{1}{2}, s_{3}=-\frac{1}{2}\right) &=-N_{h}(s-r) \sin \left(\theta_{h} / 2\right), \\
H_{2}\left(s_{2}=-\frac{1}{2}, s_{3}=\frac{1}{2}\right) &=N_{h}(s+r) \sin \left(\theta_{h} / 2\right), \\
H_{2}\left(s_{2}=-\frac{1}{2}, s_{3}=-\frac{1}{2}\right) &=N_{h}(s-r) \cos \left(\theta_{h} / 2\right),\label{eq:hadronic12H2ME2}
\end{align}
with $N_h=G_Fm_{\pi}^2\sqrt{2m_2(E_3+m_3)}$. Using the two-body phase space and squared matrix elements, we obtain the spin-averaged decay width
\begin{align}
\Gamma\left(B_{2} \rightarrow B_{3} B_{4}\right) &=\int_{-1}^{1} \frac{1}{2 m_{2}}(2 \pi)^{4} \frac{1}{(2 \pi)^{5}} \frac{\left|\vec{p}_{3}\right| d \cos \theta_{h}}{4 m_{2}} \frac{1}{2} \overline{\left|\mathcal{M}\left(B_{2} \rightarrow B_{3} B_{4}\right)\right|^{2}} \nonumber\\
&=\frac{N_{h}^{2} \sqrt{\lambda\left(m_{2}, m_{3}, m_{4}\right)}}{16 \pi m_{2}^{3}}\left(|s|^{2}+|r|^{2}\right).\end{align}

\subsection{Hadronic part $H_2$ with spin-3/2 intermediate state} \label{sec:hadronH2with32}

The transition for a spin-3/2 baryon $\Omega^-$ decaying into a spin-1/2 baryon $\Lambda$ and a spin-0 pseudoscalar meson $K^-$ is parametrized as
\begin{align}
i\mathcal{M}=i\frac{g_{h}}{m_4}\bar{u}_{B'_3}(p_3,s_3)(A-B\gamma_5)U^{\alpha}_{B'_2}(p_2,s_2)p_{4\alpha}.
\end{align}

Using the spinors and vectorial spinors, we can obtain the following terms:
\begin{align}
H_{2}\left(s_{2}=\frac{3}{2}, s_{3}=\frac{1}{2}\right) &=N_{h}^{\prime}\left|\vec{p}_{3}\right|(s+r) \sin \theta_{h} \cos \frac{\theta_{h}}{2}, \label{eq:hadronic32H2ME1}\\
H_{2}\left(s_{2}=\frac{3}{2}, s_{3}=-\frac{1}{2}\right) &=-N_{h}^{\prime}\left|\vec{p}_{3}\right|(s-r) \sin \theta_{h} \sin \frac{\theta_{h}}{2},  \\
H_{2}\left(S_{\Omega}=-\frac{3}{2}, s_{3}=\frac{1}{2}\right) &=-N_{h}^{\prime}\left|\vec{p}_{3}\right|(s+r) \sin \theta_{h} \sin \frac{\theta_{h}}{2},  \\
H_{2}\left(s_{2}=-\frac{3}{2}, s_{3}=-\frac{1}{2}\right) &=-N_{h}^{\prime}\left|\vec{p}_{3}\right|(s-r) \sin \theta_{h} \cos \frac{\theta_{h}}{2}, \\
H_{2}\left(s_{2}=\frac{1}{2}, s_{3}=\frac{1}{2}\right) &=\frac{\sqrt{3}}{3} N_{h}^{\prime}\left|\vec{p}_{3}\right|(s+r) \cos \frac{\theta_{h}}{2}\left(1-3 \cos \theta_{h}\right), \\
H_{2}\left(s_{2}=\frac{1}{2}, s_{3}=-\frac{1}{2}\right) &=\frac{\sqrt{3}}{3} N_{h}^{\prime}\left|\vec{p}_{3}\right|(s-r) \sin \frac{\theta_{h}}{2}\left(1+3 \cos \theta_{h}\right), \\
H_{2}\left(s_{2}=-\frac{1}{2}, s_{3}=\frac{1}{2}\right) &=-\frac{\sqrt{3}}{3} N_{h}^{\prime}\left|\vec{p}_{3}\right|(s+r) \sin \frac{\theta_{h}}{2}\left(1+3 \cos \theta_{h}\right), \\
H_{2}\left(s_{2}=-\frac{1}{2}, s_{3}=-\frac{1}{2}\right) &=\frac{\sqrt{3}}{3} N_{h}^{\prime}\left|\vec{p}_{3}\right|(s-r) \cos \frac{\theta_{h}}{2}\left(1-3 \cos \theta_{h}\right), \label{eq:hadronic32H2ME2}
\end{align}
where $N_h'=g_h\sqrt{(E_3+m_3)m_2}/m_4$. The spin-averaged decay width of process is given as
\begin{align}
\Gamma\left(B_{2}^{\prime} \rightarrow B_{3}^{\prime} B_{4}^{\prime}\right) &=\int_{-1}^{1} \frac{1}{2 m_{2}}(2 \pi)^{4} \frac{1}{(2 \pi)^{5}} \frac{\left|\vec{p}_{3}\right| d \cos \theta_{h}}{4 m_{2}} \frac{1}{4} \overline{\left|\mathcal{M}\left(B_{2}^{\prime} \rightarrow B_{3}^{\prime} B_{4}^{\prime}\right)\right|^{2}} \nonumber\\
&=\frac{N_{h}^{\prime 2}\left|\vec{p}_{3}\right|^{3}}{12 \pi m_{2}^{2}}\left(|s|^{2}+|r|^{2}\right).
\end{align}

%%%%%%%%%%%%%%%%%%%%%%%%%%%%%%%%%%%%%%%%%%%%%%%%%%%
%%%%%%%%%%%%%%%%%%%%%%%%%%%%%%%%%%%%%%%%%%%%%%%%%%%
%%%%%%%%%%%%%%%%%%%%%%%%%%%%%%%%%%%%%%%%%%%%%%%%%%%

\section{Theoretical results of differential decay width}

\subsection{Differential decay width of $(\Lambda_c^+, \Xi_c^0)$ }

Combining the leptonic matrix elements in Eq.(\ref{eq:leptonicME1})-(\ref{eq:leptonicME2}) and hadronic ones $H_1$ in Eq.(\ref{eq:hadronicH1ME1})-(\ref{eq:hadronicH1ME2}), together with the three-body phase space in Eq.(\ref{eq:threebodyphasespace}), we can obtain the three-body decay width of $B_1\to B_2\ell^+\nu_{\ell}$ processes, which $B_1=(\Lambda_c^+, \Xi_c^0)$ and corresponding $B_2=(\Lambda, \Xi^-)$:
\begin{align}
\frac{d \Gamma\left(B_{1} \rightarrow B_{2} l^{+} \nu_{\ell}\right)}{d \cos \theta_{\ell} d q^{2}} &=\frac{q^{2}\sqrt{\lambda\left(m_{1}, m_{2}, q\right)}}{1024 \pi^{3} m_{1}^{3}}\left(1-\hat{m}_{\ell}\right)^{2}\frac{G_{F}^{2}}{2}\left|V_{\mathrm{cs}}\right|^{2}  \nonumber\\
&\left\{\left|H_{\frac{1}{2}, 1}\right|^{2}\left[\left(1-\cos \theta_{\ell}\right)^{2}+\hat{m}_{\ell}^{2} \sin ^{2} \theta_{\ell}\right]\right. \nonumber\\
&+\left|H_{-\frac{1}{2},-1}\right|^{2}\left[\left(1+\cos \theta_{\ell}\right)^{2}+\hat{m}_{\ell}^{2} \sin ^{2} \theta_{\ell}\right] \nonumber\\
&+\left(\left|H_{\frac{1}{2}, 0}\right|^{2}+\left|H_{-\frac{1}{2}, 0}\right|^{2}\right) 2\left(\sin ^{2} \theta_{\ell}+\hat{m}_{\ell} \cos ^{2} \theta_{\ell}\right) \nonumber\\
&+\left(\left|H_{\frac{1}{2}, t}\right|^{2}+\left|H_{-\frac{1}{2}, t}\right|^{2}\right) 2 \hat{m}_{\ell}^{2}\nonumber\\
&\left.+2\left(\left|H_{\frac{1}{2}, 0}H_{\frac{1}{2}, t}^{*}\right|+\left|H_{-\frac{1}{2}, 0}H_{-\frac{1}{2}, t}^{*}\right|\right) 2 \hat{m}_{\ell}^{2}\cos\theta_{l}\right\}.
\end{align}

Besides, taking the hadronic matrix elements $H_2$ from Eq.(\ref{eq:hadronic12H2ME1})-(\ref{eq:hadronic12H2ME2}) into account, we can obtain the four-body differential decay width for $B_1\to B_3B_4\ell^+\nu_{\ell}$ defined in Eq.(\ref{eq:general4bodydecaywidth}):
\begin{align}
\frac{d \Gamma\left(B_{1} \rightarrow B_{3} B_{4} l^{+} \nu_{\ell}\right)}{d \cos \theta_{\ell} d \cos \theta_{h} d \phi d q^{2}}=&\frac{G_{F}^{2}\left|V_{\mathrm{cs}}\right|^{2} q^{2}\left(1-\hat{m}_{\ell}^{2}\right) \sqrt{\lambda\left(m_{1}, m_{2}, q\right)}}{(2 \pi)^{4} 256 m_{1}^{3}} \mathcal{B}\left(B_{2} \rightarrow B_{3} B_{4}\right) \nonumber\\
\times &\left\{\left|H_{\frac{1}{2}, 1}\right|^{2}\left(1+\alpha \cos \theta_{h} \right)\left[\hat{m}_{\ell}^{2} \sin ^{2} \theta_{\ell}+\left(1-\cos \theta_{\ell}\right)^{2}\right]\right. \nonumber\\
&+\left|H_{-\frac{1}{2},-1}\right|^{2}\left(1-\alpha \cos \theta_{h} \right)\left[\hat{m}_{\ell}^{2} \sin ^{2} \theta_{\ell}+\left(1+\cos \theta_{\ell}\right)^{2}\right] \nonumber\\
&+2\left|H_{\frac{1}{2}, 0}\right|^{2}\left(1+\alpha \cos \theta_{h} \right)\left[\hat{m}_{\ell}^{2} \cos ^{2} \theta_{\ell}+\sin ^{2} \theta_{\ell}\right] \nonumber\\
&+2\left|H_{-\frac{1}{2}, 0}\right|^{2}\left(1-\alpha \cos \theta_{h} \right)\left[\hat{m}_{\ell}^{2} \cos ^{2} \theta_{\ell}+\sin ^{2} \theta_{\ell}\right] \nonumber\\
&+2\left|H_{\frac{1}{2}, t}\right|^{2}\left(1+\alpha \cos \theta_{h} \right) \hat{m}_{\ell}^{2} \nonumber\\
&+2\left|H_{-\frac{1}{2}, t}\right|^{2}\left(1-\alpha \cos \theta_{h} \right) \hat{m}_{\ell}^{2} \nonumber\\
&+2 \sqrt{2} \alpha \sin \theta_{h} \sin \theta_{\ell} \cos \phi\left|H_{\frac{1}{2}, 1} H_{-\frac{1}{2}, 0}^{*}\right|\left(1-\cos \theta_{\ell}+\hat{m}_{\ell}^{2} \cos \theta_{\ell}\right) \nonumber\\
&+2 \sqrt{2} \alpha \sin \theta_{h} \sin \theta_{\ell} \cos \phi\left|H_{-\frac{1}{2},-1} H_{\frac{1}{2}, 0}^{*}\right|\left(1+\cos \theta_{\ell}-\hat{m}_{\ell}^{2} \cos \theta_{\ell}\right) \nonumber\\
&+2 \sqrt{2} \alpha \sin \theta_{h} \sin \theta_{\ell} \cos \phi \mid H_{\frac{1}{2}, 1} H_{-\frac{1}{2}, t}^{*} \hat{m}_{\ell}^{2}\nonumber \\
&-2 \sqrt{2} \alpha \sin \theta_{h} \sin \theta_{\ell} \cos \phi\left|H_{-\frac{1}{2},-1} H_{\frac{1}{2}, t}^{*}\right| \hat{m}_{\ell}^{2} \nonumber\\
&+4 \hat{m}_{\ell}^{2} \cos \theta_{\ell}\left|H_{-\frac{1}{2}, 0} H_{-\frac{1}{2}, t}^{*}\right|\left(1-\alpha \cos \theta_{h} \right) \nonumber\\
&\left.+4 \hat{m}_{\ell}^{2} \cos \theta_{\ell}\left|H_{\frac{1}{2}, 0} H_{\frac{1}{2}, t}^{*}\right|\left(1+\alpha \cos \theta_{h} \right)\right\}.
\end{align}

By integrating out the parameters step by step, we can get the results of $q^2$-, $\theta_l$-, $\theta_h$- and $\phi$-dependence form of total decay width respectively:
\begin{itemize}
	\item $q^2$-dependence:
\begin{align}
\frac{d \Gamma}{d q^{2}}=& \frac{\pi G_{F}^{2}\left|V_{\mathrm{cs}}\right|^{2} q^{2}\left(1-\hat{m}_{\ell}^{2}\right) \sqrt{\lambda\left(m_{1}, m_{2}, q\right)}}{(2 \pi)^{4} 48 m_{1}^{3}} \mathcal{B}\left(B_{2} \rightarrow B_{3} B_{4}\right)  \nonumber\\
& \times\left[\left(2+\hat{m}_{\ell}^{2}\right)\left(\left|H_{-\frac{1}{2},-1}\right|^{2}+\left|H_{\frac{1}{2}, 1}\right|^{2}+\left|H_{-\frac{1}{2}, 0}\right|^{2}+\left|H_{\frac{1}{2}, 0}\right|^{2}\right)+3 \hat{m}_{\ell}^{2}\left(\left|H_{-\frac{1}{2}, t}\right|^{2}+\left|H_{\frac{1}{2}, t}\right|^{2}\right)\right],\label{halfq2}
\end{align}

\item $\theta_l$-dependence:
\begin{align}
	\frac{d \Gamma}{d \cos \theta_{\ell}}=\int_{q_{\min }^{2}}^{q_{\max }^{2}} d q^{2} \frac{G_{F}^{2}\left|V_{\mathrm{cs}}\right|^{2} q^{2}\left(1-\hat{m}_{\ell}^{2}\right) \sqrt{\lambda\left(m_{1}, m_{2}, q\right)}}{(2 \pi)^{4} 256 m_{1}^{3}} \mathcal{B}\left(B_{2} \rightarrow B_{3} B_{4}\right)\left(A_{\ell}+B_{\ell} \cos \theta_{\ell}+C_{\ell} \cos 2 \theta_{\ell}\right),
\end{align}
with
\begin{align}
A_{\ell}=&2 \pi\left[\left(3+\hat{m}_{\ell}^{2}\right)\left(\left|H_{-\frac{1}{2},-1}\right|^{2}+\left|H_{\frac{1}{2}, 1}\right|^{2}\right)+2\left(1+\hat{m}_{\ell}^{2}\right)\left(\left|H_{-\frac{1}{2}, 0}\right|^{2}+\left|H_{\frac{1}{2}, 0}\right|^{2}\right)\right. \nonumber\\
&\left.\quad+4 \hat{m}_{\ell}^{2}\left(\left|H_{-\frac{1}{2}, t}\right|^{2}+\left|H_{\frac{1}{2}, t}\right|^{2}\right)\right], \\
B_{\ell}=&8 \pi\left[\left|H_{-\frac{1}{2},-1}\right|^{2}-\left|H_{\frac{1}{2}, 1}\right|^{2}+2 \hat{m}_{\ell}^{2}\left(\left|H_{-\frac{1}{2}, 0} H_{-\frac{1}{2}, t}^{*}\right|+\left|H_{\frac{1}{2}, 0} H_{\frac{1}{2}, t}^{*}\right|\right)\right], \\
C_{\ell}=&2 \pi\left(1-\hat{m}_{\ell}^{2}\right)\left[\left|H_{-\frac{1}{2},-1}\right|^{2}+\left|H_{\frac{1}{2}, 1}\right|^{2}-2\left(\left|H_{-\frac{1}{2}, 0}\right|^{2}+\left|H_{\frac{1}{2}, 0}\right|^{2}\right)\right],
\end{align}

\item $\theta_h$-dependence:
\begin{align}
	\frac{d \Gamma}{d \cos \theta_{h}}=\int_{q_{\min }^{2}}^{q_{\max }^{2}} d q^{2} \frac{G_{F}^{2}\left|V_{\mathrm{cs}}\right|^{2} q^{2}\left(1-\hat{m}_{\ell}^{2}\right) \sqrt{\lambda\left(m_{1}, m_{2}, q\right)}}{(2 \pi)^{4} 256 m_{1}^{3}} \mathcal{B}\left(B_{2} \rightarrow B_{3} B_{4}\right)\left(A_{h}+B_{h} \cos \theta_{h} \right),
\end{align}
with
\begin{align}
A_{h}=&\frac{8 \pi}{3}\left[\left(2+\hat{m}_{\ell}^{2}\right)\left(\left|H_{-\frac{1}{2},-1}\right|^{2}+\left|H_{\frac{1}{2}, 1}\right|^{2}+\left|H_{-\frac{1}{2}, 0}\right|^{2}+\left|H_{\frac{1}{2}, 0}\right|^{2}\right)+3 \hat{m}_{\ell}^{2}\left(\left|H_{-\frac{1}{2}, t}\right|^{2}+\left|H_{\frac{1}{2}, t}\right|^{2}\right)\right], \\
B_{h}=&\frac{8 \pi}{3} \alpha\left[\left(2+\hat{m}_{\ell}^{2}\right)\left(-\left|H_{-\frac{1}{2},-1}\right|^{2}+\left|H_{\frac{1}{2}, 1}\right|^{2}-\left|H_{-\frac{1}{2}, 0}\right|^{2}+\left|H_{\frac{1}{2}, 0}\right|^{2}\right)+3 \hat{m}_{\ell}^{2}\left(-\left|H_{-\frac{1}{2}, t}\right|^{2}+\left|H_{\frac{1}{2}, t}\right|^{2}\right)\right],
\end{align}

\item $\phi$-dependence:
\begin{align}
	\frac{d \Gamma}{d \phi}=\int_{q_{\min }^{2}}^{q_{\max }^{2}} d q^{2} \frac{G_{F}^{2}\left|V_{\mathrm{cs}}\right|^{2} q^{2}\left(1-\hat{m}_{\ell}^{2}\right) \sqrt{\lambda\left(m_{1}, m_{2}, q\right)}}{(2 \pi)^{4} 256 m_{1}^{3}} \mathcal{B}\left(B_{2} \rightarrow B_{3} B_{4}\right)\left(A_{\phi}+B_{\phi} \cos \phi \theta_{h} \right),\label{halfphi}
\end{align}
with
\begin{align}
A_{\phi}=&\frac{8}{3}\left[\left(2+\hat{m}_{\ell}^{2}\right)\left(\left|H_{-\frac{1}{2},-1}\right|^{2}+\left|H_{\frac{1}{2}, 1}\right|^{2}+\left|H_{-\frac{1}{2}, 0}\right|^{2}+\left|H_{\frac{1}{2}, 0}\right|^{2}\right)+3 \hat{m}_{\ell}^{2}\left(\left|H_{-\frac{1}{2}, t}\right|^{2}+\left|H_{\frac{1}{2}, t}\right|^{2}\right)\right], \\
B_{\phi}=&\frac{\pi^{2}}{\sqrt{2}} \alpha\left[\left|H_{\frac{1}{2}, 1} H_{-\frac{1}{2}, 0}^{*}\right|+\left|H_{-\frac{1}{2},-1} H_{\frac{1}{2}, 0}^{*}\right|+\hat{m}_{\ell}^{2}\left(\left|H_{\frac{1}{2}, 1} H_{-\frac{1}{2}, t}^{*}\right|-\left|H_{-\frac{1}{2},-1} H_{\frac{1}{2}, t}^{*}\right|\right)\right].
\end{align}

\end{itemize}

%%%%%%%%%%%%%%%%%%%%%%%%%%%%%%%%%%%%%%%
%%%%%%%%%%%%%%%%%%%%%%%%%%%%%%%%%%%%%%%
%%%%%%%%%%%%%%%%%%%%%%%%%%%%%%%%%%%%%%%

\subsection{Differential decay width of $\Omega_c^0$}

Using the leptonic matrix elements in Eq.(\ref{eq:leptonicME1})-(\ref{eq:leptonicME2}) and hadron ones $H_1$ in Eq.(\ref{eq:hadronic32H1start})-(\ref{eq:hadronic32H1end}), together with the three-body phase space in Eq.(\ref{eq:threebodyphasespace}), we can obtain the three-body decay width from spin-1/2  charmed baryon $\Omega_c^0$ to spin-3/2 baryon $\Omega^-$ and leptons:
\begin{align}
\frac{d \Gamma\left(\Omega_c^0 \rightarrow \Omega^- l^{+} \nu_{\ell}\right) }{d \cos \theta_{\ell} d q^{2}}&=\frac{q^{2}\sqrt{\lambda\left(m_{\Omega_c}, m_{\Omega}, q\right)}}{1024 \pi^{3} m_{\Omega_c}^{3}}\left(1-\hat{m}_{\ell}^{2}\right)^{2} \frac{G_{F}^{2}}{2}\left|V_{\mathrm{cs}}\right|^{2}   \nonumber\\
&\left\{\left(\left|H_{\frac{3}{2}, 1}\right|^{2}+\left|H_{\frac{1}{2}, 1}\right|^{2}\right)\left[\left(1-\cos \theta_{\ell}\right)^{2}+\hat{m}_{\ell}^{2} \sin ^{2} \theta_{\ell}\right]\right. \nonumber\\
&+\left(\left|H_{-\frac{3}{2},-1}\right|^{2}+\left|H_{-\frac{1}{2}, -1}\right|^{2} \right)\left[\left(1+\cos \theta_{\ell}\right)^{2}+\hat{m}_{\ell}^{2} \sin ^{2} \theta_{\ell}\right] \nonumber\\
&+\left(\left|H_{\frac{1}{2}, 0}\right|^{2}+\left|H_{-\frac{1}{2}, 0}\right|^{2}\right) 2\left(\sin ^{2} \theta_{\ell}+\hat{m}_{\ell} \cos ^{2} \theta_{\ell}\right) \nonumber\\
&+\left(\left|H_{\frac{1}{2}, t}\right|^{2}+\left|H_{-\frac{1}{2}, t}\right|^{2}\right) 2 \hat{m}_{\ell}^{2} \nonumber\\
&\left.+2\left(\left|H_{\frac{1}{2}, 0} H_{\frac{1}{2}, t} \right|^{*}+\left|H_{-\frac{1}{2}, 0}H_{-\frac{1}{2}, t} \right|^{*}\right)\left(2\hat{m}^{2}_{\ell}\cos\theta_{l} \right) \right\}.
\end{align}

Bring the results of leptonic part in Eq.(\ref{eq:leptonicME1})-(\ref{eq:leptonicME2}), hadronic part $H_1$ in Eq.(\ref{eq:hadronic32H1start})-(\ref{eq:hadronic32H1end}) and $H_2$ in Eq.(\ref{eq:hadronic32H2ME1})-(\ref{eq:hadronic32H2ME2}) together, the differential decay width of four-body process $\Omega_c^0\to\Lambda K^-\ell^+\nu_l$ can be expressed as
\begin{align}
\label{eq:omegafourbody}
 \frac{d \Gamma \left(\Omega_c^0\to\Lambda K^-\ell^+\nu_l\right)}{d q^{2} d \cos \theta_{h} d \cos \theta_{\ell} d \phi} 
=& \frac{3 G_{F}^{2}\left|V_{\mathrm{cs}}\right|^{2} q^{2}\left(1-\hat{m}_{\ell}^{2}\right)^{2} \sqrt{\lambda\left(m_{\Omega_{c}}, m_{\Omega}, q\right)} \mathcal{B}(\Omega^- \rightarrow \Lambda K^-)}{512(2 \pi)^{4} m_{\Omega_{c}}^{3}} \nonumber\\
\times &\left\{\left|H_{-\frac{3}{2},-1}\right|^{2}\left[\hat{m}_{\ell}^{2} \sin ^{2} \theta_{\ell}+\left(1+\cos \theta_{\ell}\right)^{2}\right] \sin ^{2} \theta_{h} \left(1-\alpha \cos \theta_{h} \right)\right.\nonumber\\
&+\left|H_{\frac{3}{2}, 1}\right|^{2}\left[\hat{m}_{\ell}^{2} \sin ^{2} \theta_{\ell}+\left(1-\cos \theta_{\ell}\right)^{2}\right] \sin ^{2} \theta_{h} \left(1+\alpha \cos \theta_{h} \right) \nonumber\\
&+\frac{1}{12}\left|H_{\frac{1}{2}, 1}\right|^{2}\left[\hat{m}_{\ell}^{2} \sin ^{2} \theta_{\ell}+\left(1-\cos \theta_{\ell}\right)^{2}\right]\left[2\left(5+3 \cos 2 \theta_{h} \right)+\alpha\left(7 \cos \theta_{h} +9 \cos 3 \theta_{h} \right)\right] \nonumber\\
&+\frac{1}{12}\left|H_{-\frac{1}{2},-1}\right|^{2}\left[\hat{m}_{\ell}^{2} \sin ^{2} \theta_{\ell}+\left(1+\cos \theta_{\ell}\right)^{2}\right]\left[2\left(5+3 \cos 2 \theta_{h} \right)-\alpha\left(7 \cos \theta_{h} +9 \cos 3 \theta_{h} \right)\right] \nonumber\\
&+\frac{1}{6}\left|H_{-\frac{1}{2}, 0}\right|^{2}\left[\hat{m}_{\ell}^{2} \cos ^{2} \theta_{\ell}+\sin ^{2} \theta_{\ell}\right]\left[2\left(5+3 \cos 2 \theta_{h} \right)-\alpha\left(7 \cos \theta_{h} +9 \cos 3 \theta_{h} \right)\right] \nonumber\\
&+\frac{1}{6}\left|H_{\frac{1}{2}, 0}\right|^{2}\left[\hat{m}_{\ell}^{2} \cos ^{2} \theta_{\ell}+\sin ^{2} \theta_{\ell}\right]\left[2\left(5+3 \cos 2 \theta_{h} \right)+\alpha\left(7 \cos \theta_{h} +9 \cos 3 \theta_{h} \right)\right] \nonumber\\
&+\frac{1}{6}\left|H_{-\frac{1}{2}, t}\right|^{2} \hat{m}_{\ell}^{2}\left[2\left(5+3 \cos 2 \theta_{h} \right)-\alpha\left(7 \cos \theta_{h} +9 \cos 3 \theta_{h} \right)\right] \nonumber\\
&+\frac{1}{6}\left|H_{\frac{1}{2}, t}\right|^{2} \hat{m}_{\ell}^{2}\left[2\left(5+3 \cos 2 \theta_{h} \right)+\alpha\left(7 \cos \theta_{h} +9 \cos 3 \theta_{h} \right)\right] \nonumber\\
&+\frac{1}{3} \mid H_{-\frac{1}{2}, 0} H_{-\frac{1}{2}, t}^{*} \hat{m}_{\ell}^{2} \cos \theta_{\ell}\left[2\left(5+3 \cos 2 \theta_{h} \right)-\alpha\left(7 \cos \theta_{h} +9 \cos 3 \theta_{h} \right)\right]  \nonumber\\
&+\frac{1}{3}\left|H_{\frac{1}{2}, 0} H_{\frac{1}{2}, t}\right| \hat{m}_{\ell}^{2} \cos \theta_{\ell}\left[2\left(5+3 \cos 2 \theta_{h} \right)+\alpha\left(7 \cos \theta_{h} +9 \cos 3 \theta_{h} \right)\right]  \nonumber\\
&+\frac{2 \sqrt{3}}{3}\left|H_{-\frac{3}{2},-1} H_{\frac{1}{2}, 1}^{*}\right|\left(\hat{m}_{\ell}^{2}-1\right) \sin ^{2} \theta_{\ell} \cos 2 \phi \sin ^{2} \theta_{h} \left(1-3 \alpha \cos \theta_{h} \right)  \nonumber\\
&+\frac{2 \sqrt{3}}{3}\left|H_{\frac{3}{2}, 1} H_{-\frac{1}{2},-1}^{*}\right|\left(\hat{m}_{\ell}^{2}-1\right) \sin ^{2} \theta_{\ell} \cos 2 \phi \sin ^{2} \theta_{h} \left(1+3 \alpha \cos \theta_{h} \right)  \nonumber\\
&+\frac{\sqrt{6}}{6}\left|H_{-\frac{3}{2},-1} H_{-\frac{1}{2}, 0}^{*}\right| \sin \theta_{\ell} \cos \phi\left\{\left(\sin \theta_{h} -3 \sin 3 \theta_{h} \right)\left[\hat{m}_{\ell}^{2} \cos \theta_{\ell}+\alpha\left(1+\cos \theta_{\ell}\right)\right]\right.  \nonumber\\
&\qquad\qquad\qquad\qquad\qquad\qquad\qquad\qquad\left.+4 \sin 2 \theta_{h} \left(\alpha \hat{m}_{\ell}^{2} \cos \theta_{\ell}+1+\cos \theta_{\ell}\right)\right\}  \nonumber\\
&-\frac{\sqrt{6}}{6}\left|H_{\frac{3}{2}, 1} H_{\frac{1}{2}, 0}^{*}\right|\left[\hat{m}_{\ell}^{2} \cos \theta_{\ell}+\left(1-\cos \theta_{\ell}\right)\right] \sin \theta_{\ell} \cos \phi\left[4 \sin 2 \theta_{h} -\alpha\left(\sin \theta_{h} -3 \cos 3 \theta_{h} \right)\right]  \nonumber\\
&+\frac{\sqrt{6}}{6}\left|H_{-\frac{3}{2},-1} H_{-\frac{1}{2}, t}^{*}\right| \hat{m}_{\ell}^{2} \sin \theta_{\ell} \cos \phi\left[\left(\sin \theta_{h} -3 \sin 3 \theta_{h} \right)+4 \alpha \sin 2 \theta_{h} \right]  \nonumber\\
&-\frac{\sqrt{6}}{6}\left|H_{\frac{3}{2}, 1} H_{\frac{1}{2}, t}^{*}\right| \hat{m}_{\ell}^{2} \sin \theta_{\ell} \cos \phi\left[4 \sin 2 \theta_{h} -\alpha\left(\sin \theta_{h} -3 \cos 3 \theta_{h} \right)\right]  \nonumber\\
&-\frac{\sqrt{2}}{6}\left|H_{\frac{1}{2}, 1} H_{-\frac{1}{2}, 0}^{*}\right|\left[\cos \theta_{\ell} \hat{m}_{\ell}^{2}-\alpha\left(1-\cos \theta_{\ell}\right)\right] \sin \theta_{\ell} \cos \phi\left(5 \sin \theta_{h} +9 \sin 3 \theta_{h} \right) \nonumber \\
&+\frac{\sqrt{2}}{6}\left|H_{-\frac{1}{2},-1} H_{\frac{1}{2}, 0}^{*}\right|\left[\left(1+\cos \theta_{\ell}\right)-\alpha \hat{m}_{\ell}^{2} \cos \theta_{\ell}\right] \sin \theta_{\ell} \cos \phi\left(5 \sin \theta_{h} +9 \sin 3 \theta_{h} \right)  \nonumber\\
&-\frac{\sqrt{2}}{6}\left|H_{\frac{1}{2}, 1} H_{-\frac{1}{2}, t}^{*}\right| \hat{m}_{\ell}^{2} \sin \theta_{\ell} \cos \phi\left(5 \sin \theta_{h} +9 \sin 3 \theta_{h} \right)  \nonumber\\
&\left.-\frac{\sqrt{2}}{6}\left|H_{-\frac{1}{2},-1} H_{\frac{1}{2}, t}^{*}\right| \hat{m}_{\ell}^{2} \sin \theta_{\ell} \cos \phi \alpha\left(5 \sin \theta_{h} +9 \sin 3 \theta_{h} \right)\right\}.
\end{align}

By integrating out the parameters step by step, we can get the results of $q^2$-, $\theta_l$-, $\theta_h$- and $\phi$-dependence form of total decay width respectively:
\begin{itemize}
	\item $q^2$-dependence:
	\begin{align}
\frac{d \Gamma}{d q^{2}}=\frac{32 \pi}{9} &\left[\left(2+\hat{m}_{\ell}^{2}\right)\left(\left|H_{-\frac{3}{2},-1}\right|^{2}+\left|H_{\frac{3}{2}, 1}\right|^{2}+\left|H_{-\frac{1}{2},-1}\right|^{2}+\left|H_{\frac{1}{2}, 1}\right|^{2}+\left|H_{-\frac{1}{2}, 0}\right|^{2}+\left|H_{\frac{1}{2}, 0}\right|^{2}\right)\right. \nonumber\\
&\left.+3 \hat{m}_{\ell}^{2}\left(\left|H_{-\frac{1}{2}, t}\right|^{2}+\left|H_{\frac{1}{2}, t}\right|^{2}\right)\right]
	\end{align}
	
	\item $\theta_l$-dependence:
	\begin{align}
		\frac{d \Gamma}{d \cos \theta_{\ell}}=\int_{q_{\mathrm{min}^2}}^{q_{\mathrm{max}^2}} dq^2
			\frac{3 G_{F}^{2}\left|V_{\mathrm{cs}}\right|^{2} q^{2}\left(1-\hat{m}_{\ell}^{2}\right)^{2} \sqrt{\lambda\left(m_{\Omega_c}, m_{\Omega}, q\right)}}{512(2 \pi)^{4} m_{\Omega_c}^{3}} \mathcal{B}\left(B_{2}^{\prime} \rightarrow B_{3}^{\prime} B_{4}^{\prime}\right)\left(A_{\ell}^{\prime}+B_{\ell}^{\prime} \cos \theta_{\ell}+C_{\ell}^{\prime} \cos 2 \theta_{\ell}\right),
	\end{align}
	with
	\begin{align}
A_{\ell}^{\prime}=& \frac{4 \pi}{3}\left[\left(3+\hat{m}_{\ell}^{2}\right)\left(\left|H_{\frac{3}{2}, 1}\right|^{2}+\left|H_{-\frac{3}{2},-1}\right|^{2}+\left|H_{\frac{1}{2}, 1}\right|^{2}+\left|H_{-\frac{1}{2},-1}\right|^{2}\right)\right.\nonumber\\
&\left.+2\left(1+\hat{m}_{\ell}^{2}\right)\left(\left|H_{\frac{1}{2}, 0}\right|^{2}+\left|H_{-\frac{1}{2}, 0}\right|^{2}\right)+4 \hat{m}_{\ell}^{2}\left(\left|H_{\frac{1}{2}, t}\right|^{2}+\left|H_{-\frac{1}{2}, t}\right|^{2}\right)\right], \\
B_{\ell}^{\prime}=&\frac{16 \pi}{3} \left[\left(\left|H_{-\frac{3}{2},-1}\right|^{2}-\left|H_{\frac{3}{2}, 1}\right|^{2}+\left|H_{-\frac{1}{2},-1}\right|^{2}-\left|H_{\frac{1}{2}, 1}\right|^{2}\right)\right.\nonumber\\
&\left.+2 \hat{m}_{\ell}^{2}\left(\left|H_{-\frac{1}{2}, 0} H_{-\frac{1}{2}, t}^{*}\right|+\left|H_{\frac{1}{2}, 0} H_{\frac{1}{2}, t \mid}^{*}\right|\right)\right], \\
C_{\ell}^{\prime}=&\frac{4 \pi}{3}(1\left.-\hat{m}_{\ell}^{2}\right)\left[\left(\left|H_{\frac{3}{2}, 1}\right|^{2}+\left|H_{-\frac{3}{2},-1}\right|^{2}+\left|H_{\frac{1}{2}, 1}\right|^{2}+\left|H_{-\frac{1}{2},-1}\right|^{2}\right)\right. \nonumber\\
&\left.-2\left(\left|H_{\frac{1}{2}, 0}\right|^{2}+\left|H_{-\frac{1}{2}, 0}\right|^{2}\right)\right],
	\end{align}
	
	\item $\theta_h$-dependence:
	\begin{align}
\frac{d \Gamma}{d \cos \theta_{h}}=\int_{q_{\min }^{2}}^{q_{\max }^{2}} d q^{2} \frac{3 G_{F}^{2}\left|V_{\mathrm{cs}}\right|^{2} q^{2}\left(1-\hat{m}_{\ell}^{2}\right)^{2} \sqrt{\lambda\left(m_{\Omega_c}, m_{\Omega}, q\right)}}{512(2 \pi)^{4} m_{\Omega_c}^{3}} \mathcal{B}\left(B_{2}^{\prime} \rightarrow B_{3}^{\prime} B_{4}^{\prime}\right) \nonumber\\
\times\left(A_{h}^{\prime}+B_{h}^{\prime} \cos \theta_{h}+C_{h}^{\prime} \cos 2 \theta_{h}+D_{h}^{\prime} \cos 3 \theta_{h}\right),
	\end{align}
	with
	\begin{align}
A_{h}^{\prime}=\frac{4 \pi}{9}&\left\{\left(2+\hat{m}_{\ell}^{2}\right)\left[3\left(\left|H_{\frac{3}{2}, 1}\right|^{2}+\left|H_{-\frac{3}{2},-1}\right|^{2}\right)+5\left(\left|H_{\frac{1}{2}, 1}\right|^{2}+\left|H_{-\frac{1}{2},-1}\right|^{2}+\left|H_{\frac{1}{2}, 0}\right|^{2}+\left|H_{-\frac{1}{2}, 0}\right|^{2}\right)\right]\right. \nonumber\\
&\left.+15 \hat{m}_{\ell}^{2}\left(\left|H_{\frac{1}{2}, t}\right|^{2}+\left|H_{-\frac{1}{2}, t}\right|^{2}\right)\right\}, \\
%%%
B_{h}^{\prime}=\frac{2 \pi \alpha}{9} &\left\{\left(2+\hat{m}_{\ell}^{2}\right)\left[3\left(\left|H_{\frac{3}{2}, 1}\right|^{2}-\left|H_{-\frac{3}{2},-1}\right|^{2}\right)+7\left(\left|H_{\frac{1}{2}, 1}\right|^{2}-\left|H_{-\frac{1}{2},-1}\right|^{2}+\left|H_{\frac{1}{2}, 0}\right|^{2}-\left|H_{-\frac{1}{2}, 0}\right|^{2}\right)\right]\right. \nonumber\\
&\left.+21 \hat{m}_{\ell}^{2}\left(\left|H_{\frac{1}{2}, t}\right|^{2}-\left|H_{-\frac{1}{2}, t}\right|^{2}\right)\right\}, \\
%%%
C_{h}^{\prime}=\frac{4 \pi}{3}&\left[\left(2+\hat{m}_{\ell}^{2}\right)\left(-\left|H_{\frac{3}{2}, 1}\right|^{2}-\left|H_{-\frac{3}{2},-1}\right|^{2}+\left|H_{\frac{1}{2}, 1}\right|^{2}+\left|H_{-\frac{1}{2},-1}\right|^{2}+\left|H_{\frac{1}{2}, 0}\right|^{2}+\left|H_{-\frac{1}{2}, 0}\right|^{2}\right)\right. \nonumber\\
&\left.+3 \hat{m}_{\ell}^{2}\left(\left|H_{\frac{1}{2}, t}\right|^{2}+\left|H_{-\frac{1}{2}, t}\right|^{2}\right)\right], \\
%%%
D_{h}^{\prime}=\frac{2 \pi \alpha}{3} &\left\{\left(2+\hat{m}_{\ell}^{2}\right)\left[\left(\left|H_{-\frac{3}{2},-1}\right|^{2}-\left|H_{\frac{3}{2}, 1}\right|^{2}\right)+3\left(\left|H_{\frac{1}{2}, 1}\right|^{2}-\left|H_{-\frac{1}{2},-1}\right|^{2}+\left|H_{\frac{1}{2}, 0}\right|^{2}-\left|H_{-\frac{1}{2}, 0}\right|^{2}\right)\right]\right.\nonumber\\
&\left.-9 \hat{m}_{\ell}^{2}\left(\left|H_{-\frac{1}{2}, t}\right|^{2}-\left|H_{\frac{1}{2}, t}\right|^{2}\right)\right\},
	\end{align}

	\item $\phi$-dependence:
	\begin{align}
		\frac{d \Gamma}{d \phi}=\int_{q_{\min }^{2}}^{q_{\max }^{2}} d q^{2} \frac{3 G_{F}^{2}\left|V_{\mathrm{cs}}\right|^{2} q^{2}\left(1-\hat{m}_{\ell}^{2}\right)^{2} \sqrt{\lambda\left(m_{\Omega_c}, m_{\Omega}, q\right)}}{512(2 \pi)^{4} m_{\Omega_c}^{3}} \mathcal{B}\left(B_{2}^{\prime} \rightarrow B_{3}^{\prime} B_{4}^{\prime}\right)\left(A_{\phi}^{\prime}+B_{\phi}^{\prime} \cos \phi+C_{\phi}^{\prime} \cos 2 \phi\right),
	\end{align}
	with
	\begin{align}
A_{\phi}^{\prime}=& \frac{16}{9}\left[\left(2+\hat{m}_{\ell}^{2}\right)\left(\left|H_{-\frac{3}{2},-1}\right|^{2}+\left|H_{\frac{3}{2}, 1}\right|^{2}+\left|H_{-\frac{1}{2},-1}\right|^{2}+\left|H_{\frac{1}{2}, 1}\right|^{2}+\left|H_{-\frac{1}{2}, 0}\right|^{2}+\left|H_{\frac{1}{2}, 0}\right|^{2}\right)\right. \nonumber\\
&\quad\qquad\left.+3 \hat{m}_{\ell}^{2}\left(\left|H_{-\frac{1}{2}, t}\right|^{2}+\left|H_{\frac{1}{2}, t}\right|^{2}\right)\right], \\
B_{\phi}^{\prime}=&\frac{\pi^{2}}{12 \sqrt{2}}\left[\sqrt{3} \alpha\left(\left|H_{-\frac{3}{2},-1} H_{-\frac{1}{2}, 0}^{*}\right|+\left|H_{\frac{3}{2}, 1} H_{\frac{1}{2}, 0}^{*}\right|\right)+\sqrt{3} \hat{m}_{\ell}^{2}\left(\left|H_{-\frac{3}{2},-1} H_{-\frac{1}{2}, t}^{*}\right|+\alpha\left|H_{\frac{3}{2}, 1} H_{\frac{1}{2}, t}^{*}\right|\right)\right. \nonumber\\
&\quad\qquad\left.+5\left(\left|H_{-\frac{1}{2},-1} H_{\frac{1}{2}, 0}^{*}\right|+\alpha\left|H_{\frac{1}{2}, 1} H_{-\frac{1}{2}, 0}^{*}\right|\right)-5 \hat{m}_{\ell}^{2}\left(\alpha\left|H_{-\frac{1}{2},-1} H_{\frac{1}{2}, t}^{*}\right|+\left|H_{\frac{1}{2}, 1} H_{-\frac{1}{2}, t}^{*}\right|\right)\right], \\
C_{\phi}^{\prime}=&-\frac{32\left(1-\hat{m}_{\ell}\right)^{2}}{9 \sqrt{3}}\left(\left|H_{-\frac{3}{2},-1} H_{\frac{1}{2}, 1}^{*}\right|+\left|H_{\frac{3}{2}, 1} H_{-\frac{1}{2},-1}^{*}\right|\right).
	\end{align}
	
\end{itemize}

\section{Numerical analysis}

\subsection{Input}

\begin{table}[!htbp]
  \centering
  \begin{tabular}{c|c|c|c|c|c|c}
  \hline
  \hline
  Baryons&$\Lambda^{+}_{c}$&$\Xi^{0}_{c}$&$\Omega^{0}_{c}$&$\Lambda$&$\Xi^{-}$ &$\Omega^{-}$ \\
  \hline
  masses(GeV)&2.2860&2.4709&2.6952&1.1156&1.3217&1.6724\\
  \hline
  lifetimes($10^{-13}$s)&2.024&1.530 & 2.680 & 2632 &1639 & 821\\
  \hline
  \end{tabular}
  \caption{The masses of initial- and final-state baryons}~\label{baryon mass}
\end{table}

In this section, all parameters used in calculation will be collected, including the baryon masses, and CKM matrix. 
In addition, the lepton mass $m_{e}$=~0.005~GeV, $m_{\mu}$=~0.1134~GeV, and  Fermi constant $G_F=1.166 \times 10^{-5}$~GeV$^{-2}$. The CKM matrix element $V_{cs}=0.973$. In the calculation of heavy baryons four-body decays, the asymmetry parameters and branch ratios are collected as \cite{Tanabashi:2018oca} 
\begin{align}
&\alpha(\Lambda^{0}\rightarrow p \pi^{-})=0.732 , ~\alpha(\Xi^{-} \rightarrow \Lambda^{0} \pi^{-})=-0.401,~\alpha(\Omega^{-} \rightarrow \Lambda^{0} k^{-})=0.0157\;,\nonumber \\
&\mathcal{B}(\Lambda^{0}\rightarrow p \pi^{-})=63.900\%,~\mathcal{B}(\Xi^{-} \rightarrow \Lambda^{0} \pi^{-})=99.887\%,~\mathcal{B}(\Omega^{-} \rightarrow \Lambda^{0} K^{-})=67.800\%\; .
\end{align}

\subsection{Numerical Results and Discussions}

In hadronic part with spin-$1/2$ intermediate state, we use the lattice QCD calculation of the $\Lambda_{c}^+\rightarrow\Lambda$ ~\cite{Meinel:2016dqj} and $\Xi_{c}^0\rightarrow \Xi^-$ \cite{Q.A.Zhang:2021} for the decay process of $\Lambda_{c}^+\rightarrow\Lambda\ell^+\nu_{\ell}$ and $\Xi_{c}^0\rightarrow\Xi^-\ell^+\nu_{\ell}$ respectively. In order to access the $q^2$-distribution, we use the modified z-expansions in the physical limits ~\cite{Bourrely:2008za}, and  the fit functions are shown as
\begin{align}
f(q^2)=\frac{1}{1-q^2/m^{2}_{\mathrm{pole}}}\sum^{n_{\mathrm{max}}}_{n=0} a^{f}_{n}[z(q^{2})]^{n}\;. 
\end{align} 
The expansion variable is defined as
\begin{align}
z(q^{2})=\frac{\sqrt{t_{+}-q^2}-\sqrt{t_{+}-t_{0}}}{\sqrt{t_{+}-q^2}+\sqrt{t_{+}-t_{0}}}\;,
\end{align}
with $t_{0}=q^{2}_{\mathrm{max}}=(m_{\Xi_{c}}-m_{\Xi})^{2}$ for $\Xi_{c}\rightarrow\Xi$ ,$t_{0}=q^{2}_{\mathrm{max}}=(m_{\Lambda_{c}}-m_{\Lambda})^{2}$ for $\Lambda_{c}\rightarrow\Lambda$, and $t_{+} =(m_{D}+m_{K})^{2}$. The pole masses in the form factors are used as $m^{f_{+},f_{\perp}}_{\mathrm{pole}}=~2.12$~GeV,$m^{f_{0}}_{\mathrm{pole}}=~2.318$~GeV,$m^{g_{+},g_{\perp}}_{\mathrm{pole}}=~2.460$~GeV, and $m^{g_{0}}_{pole}=~1.968$~GeV.

\begin{itemize}

 \item $\Lambda_{c}^+\rightarrow p\pi^-\ell^+\nu_{\ell}$ 
 
 We collect the fitted form factors parameters from \cite{Meinel:2016dqj} in Table~\ref{lambdac form factor}. The resulting SM predictions for the $\Lambda_{c}^+\rightarrow p\pi^-\ell^+\nu_{\ell}$ decay widths and branching fractions  with corresponding error estimates are listed as
 \begin{align}
 & \Lambda_{c}^+\rightarrow p \pi^{-}e^{+}\nu_{e}:~~  \Gamma=~8.061(482)\times 10^{-14} s^{-1} , ~~  \mathcal{B}=~2.48\%(15),  \\
 & \Lambda_{c}^+\rightarrow p \pi^{-}\mu^{+}\nu_{\mu}:~~ \Gamma=~8.126(462)\times 10^{-14} s^{-1}, ~~  \mathcal{B}=~2.50\%(14).  
 \end{align}
 
  \begin{table}[!h]
  \centering
  \begin{tabular}{c|c|c|c|c|c|c|c|c|c}
  \hline
  \hline
&$a^{f_{\perp}}_{0}$&$a^{f_{\perp}}_{1}$&$a^{f_{\perp}}_{2}$&$a^{f_{+}}_{0}$&$a^{f_{+}}_{1}$&$a^{f_{+}}_{2}$&$a^{f_{0}}_{0}$&$a^{f_{0}}_{1}$&$a^{f_{0}}_{2}$\\  \hline
Nominal&$1.30\pm0.06$&$-3.27\pm1.18$&$7.16\pm11.6$&$0.81\pm0.03$&$-2.89\pm0.52$&$7.82\pm4.53$&$0.77\pm0.02$&$-2.24\pm0.51$&$5.38\pm4.80$\\ \hline
&$a^{g_{\perp}}_{0}$&$a^{g_{\perp}}_{1}$&$a^{g_{\perp}}_{2}$&$a^{g_{+}}_{0}$&$a^{g_{+}}_{1}$&$a^{g_{+}}_{2}$&$a^{g_{0}}_{0}$&$a^{g_{0}}_{1}$&$a^{g_{0}}_{2}$\\  \hline
Nominal&$0.68\pm0.02$&$-1.19\pm0.35$&$6.24\pm4.89$&$0.68\pm0.02$&$-2.44\pm0.25$&$13.7\pm2.15$&$0.71\pm0.03$&$-2.86\pm0.44$&$11.8\pm2.47$\\  \hline
  \end{tabular}
  \caption{{The $\Lambda_{c}\rightarrow\Lambda$ form factors calculated on lattice \cite{Meinel:2016dqj} .}}~\label{lambdac form factor}
\end{table}

  Based on the  form factors, we predict the differential decay widths for $\Lambda_{c}^+\rightarrow p \pi^{-}\ell^{+}\nu_{\ell}$ as function of $q^{2}$ in Fig.~\ref{Lambda angular distributionA}(a). Note that the increasing errors in small $q^{2}$ region come from the uncertainties of form factors at large momentum transfer in lattice calculations. In Eq.(\ref{halfq2}-\ref{halfphi}), we show the theoretical results of $\theta_{l}-$, $\theta_{h}$- and $\phi$-dependence of total four-body decay width  with different final state leptons, in which the coefficients are functions of $q^{2}$ only. After we integrate out $q^{2}$, the angular distribution with $\cos \theta_{l}$, $\cos \theta_{h}$ and $\phi$ are shown as
 \begin{enumerate}
 \item $\Lambda_{c}^+\rightarrow p \pi^{-} e^{+}\nu_{e}$ 
\begin{align}
\frac{d\Gamma}{\Gamma d\cos\theta_{l}}&=0.4448(254)+0.1992(172)\cos\theta_{l}-0.1657(204)\cos2\theta_{l}\;, \\
\frac{d\Gamma}{\Gamma d\cos\theta_{h}}&=0.5000(299)-0.3188(210)\cos\theta_{h}\;, \\
\frac{d\Gamma}{\Gamma d\phi}&=0.1592(95)-0.0276(35)\cos\theta_{\phi}\;.
\end{align}

\item  $\Lambda_{c}^+\rightarrow p \pi^{-} \mu^{+}\nu_{\mu}$ 
 
 \begin{align}
\frac{d\Gamma}{\Gamma d\cos\theta_{l}}&=0.455(57)+0.265(28)\cos\theta_{l}-0.135(24)\cos2\theta_{l}\;, \\
\frac{d\Gamma}{\Gamma d\cos\theta_{h}}&=0.500(63)+0.162(19)\cos\theta_{h}\;, \\
\frac{d\Gamma}{\Gamma d\phi}&=0.159(20)+0.012(3)\cos\theta_{\phi}\;.
\end{align}
\end{enumerate}

\begin{figure}[htbp]
\centering
\includegraphics[width=1.0\columnwidth]{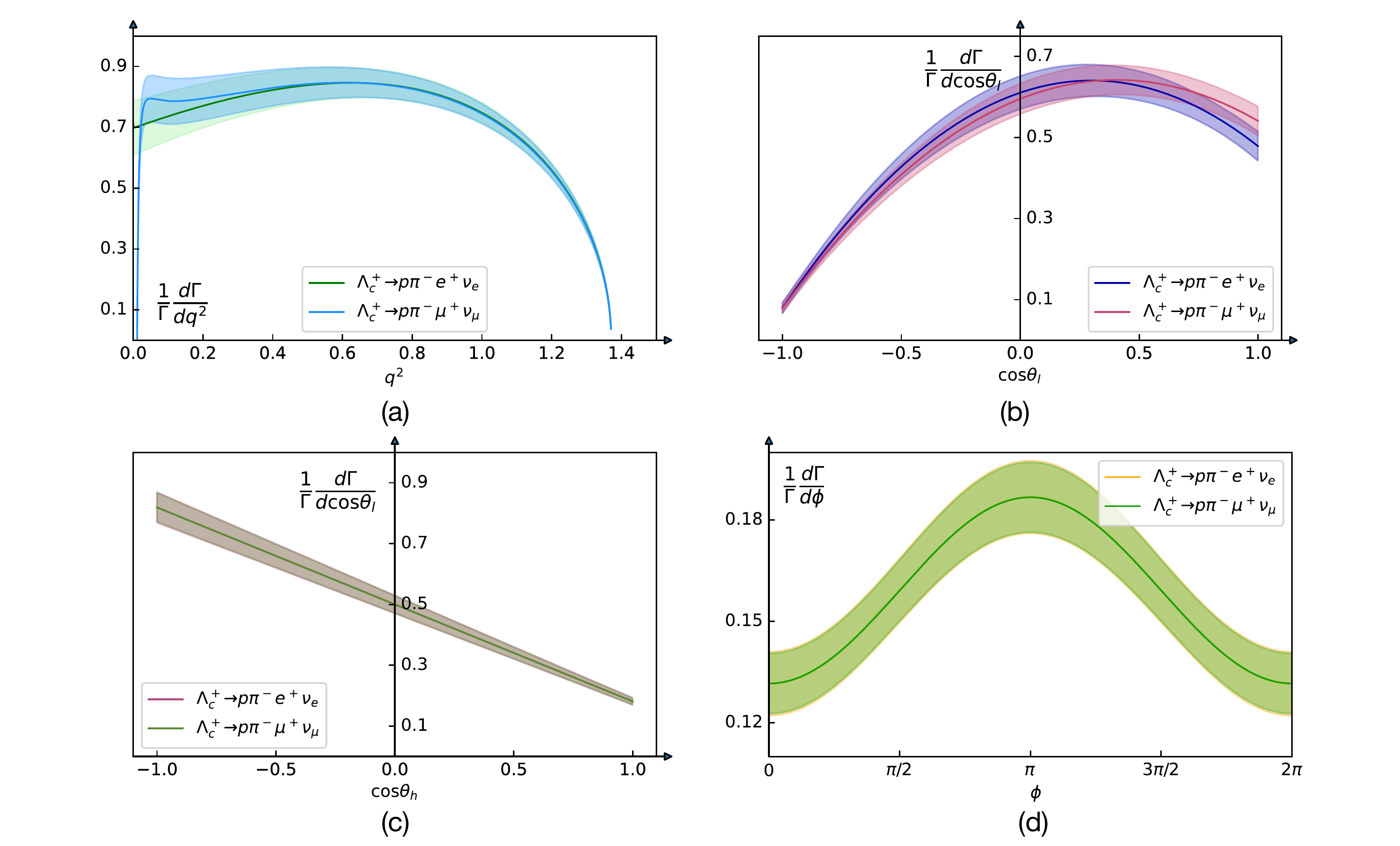}
%\caption{}
\centering

%\caption{fig2}
\caption{ Differential decay width  for $\Lambda_{c}^+\rightarrow p \pi^{-}\ell^{+}\nu_{\ell}$ as function of $q^{2}$ (a), $\cos\theta_{l}$ (b), $\cos\theta_{h}$ (c), and $\phi$ (d) respectively. It should be noted that in sub-fig.(b) and (c), the errors increasing with $\theta_{l}$ form -1 to 1 and $\theta_{h}$ from 1 to -1, as a results of the remarkable uncertainties of form factors at small $q^2$ region. }
\label{Lambda angular distributionA}
\end{figure}

 \item $\Xi_{c}^0\rightarrow \Lambda \pi^{-} \ell^+\nu_{\ell}$ 
 
 Based on the recently announced $\Xi_{c}\to\Xi$ form factors calculated by lattice QCD~\cite{Q.A.Zhang:2021},   where the  $z$-expansion parameters of helicity-based form factors collected in Table~\ref{xic form factor}, we perform the predictions for $\Xi_{c}^0\rightarrow \Lambda \pi^{-} \ell^+\nu_{\ell}$ decay widths and branching fractions  with corresponding uncertainties  
  \begin{align}
  \Xi_{c}\rightarrow \Lambda \pi^{-}e^{+}\nu_{e}:~~ & \Gamma=1.031(129)\times 10^{-13} s^{-1} ,~~  \mathcal{B}=2.40(30)\%,  \\
  \Xi_{c}\rightarrow \Lambda \pi^{-}\mu^{+}\nu_{\nu}:~~& \Gamma=1.037(131)\times 10^{-13} s^{-1}, ~~  \mathcal{B}=2.41(30)\%. 
 \end{align}

   \begin{table}[!htbp]
  \centering
  \begin{tabular}{c|c|c|c|c|c|c}
  \hline
  \hline
&$f_{\perp}$&$f_{0}$&$f_{+}$&$g_{\perp}$&$g_{0}$&$g_{+}$\\ 
\hline
$a_{0}$&$1.51\pm0.09$&$0.64\pm0.09$&$0.77\pm0.07$&$0.56\pm0.07$&$0.63\pm0.07$&$0.56\pm0.08$\\ \hline
$a_{1}$&$-1.88\pm1.21$&$-1.83\pm1.22$&$-4.09\pm1.18$&$-0.35\pm1.26$&$-1.63\pm1.36$&$0.00\pm1.38$\\ \hline
$a_{2}$&$1.71\pm0.49$&$0.56\pm0.51$&$0.35\pm0.49$&$0.15\pm0.29$&$0.15\pm0.29$&$0.14\pm0.29$ \\ \hline
  \end{tabular}
  \caption{Results for the z-expansion parameters describing  $\Xi_{c}\rightarrow\Xi$ form factors with statistical errors. }~\label{xic form factor}
\end{table}

We also predict the differential decay widths for $\Xi_{c}^0\rightarrow \Lambda \pi^{-}\ell^{+}\nu_{\ell}$ as function of $q^{2}$ in Fig.~\ref{Xi angular distributionA}, the errors in plots mainly come from the uncertainties of form factors extracted from lattice QCD, and its behaviors is same as  $\Lambda_{c}^+\rightarrow p \pi^{-}\ell^{+}\nu_{\ell}$. The results of $\theta_{l}$-, $\theta_{h}$- and $\phi$-dependence of differential decay widths with different leptonic final-states, are listed as follows
  \begin{enumerate}
  \item $\Xi_{c}^0\rightarrow \Lambda \pi^{-} e^{+}\nu_{e}$ 
\begin{align}
&\frac{d\Gamma}{\Gamma d\cos\theta_{l}}=0.4478(557)+0.2363(249)\cos\theta_{l}-0.1565(273)\cos2\theta_{l}\;,\nonumber \\
&\frac{d\Gamma}{\Gamma d\cos\theta_{h}}=0.5000(626)+0.1616(191)\cos\theta_{h}\;,\nonumber \\
&\frac{d\Gamma}{\Gamma d\phi}=0.1592(199)+0.0123(30)\cos\theta_{\phi}\;.
\end{align}

 \item $\Xi_{c}^0\rightarrow \Lambda \pi^{-} \mu^{+}\nu_{\mu}$ 
 
 \begin{align}
&\frac{d\Gamma}{\Gamma d\cos\theta_{l}}=0.4549(569)+0.2650(280)\cos\theta_{l}-0.1354(240)\cos2\theta_{l}\;,\nonumber \\
&\frac{d\Gamma}{\Gamma d\cos\theta_{h}}=0.500( 630)+0.1617(193)\cos\theta_{h}\;,\nonumber \\
&\frac{d\Gamma}{\Gamma d\phi}=0.1592(200)+0.0122(29)\cos\theta_{\phi}\;,
\end{align}
 \end{enumerate}
 and corresponding plots of angular distributions are collected in Fig.\ref{Xi angular distributionA}.
 
\begin{figure}[htbp]
\centering
\includegraphics[width=1.0\columnwidth]{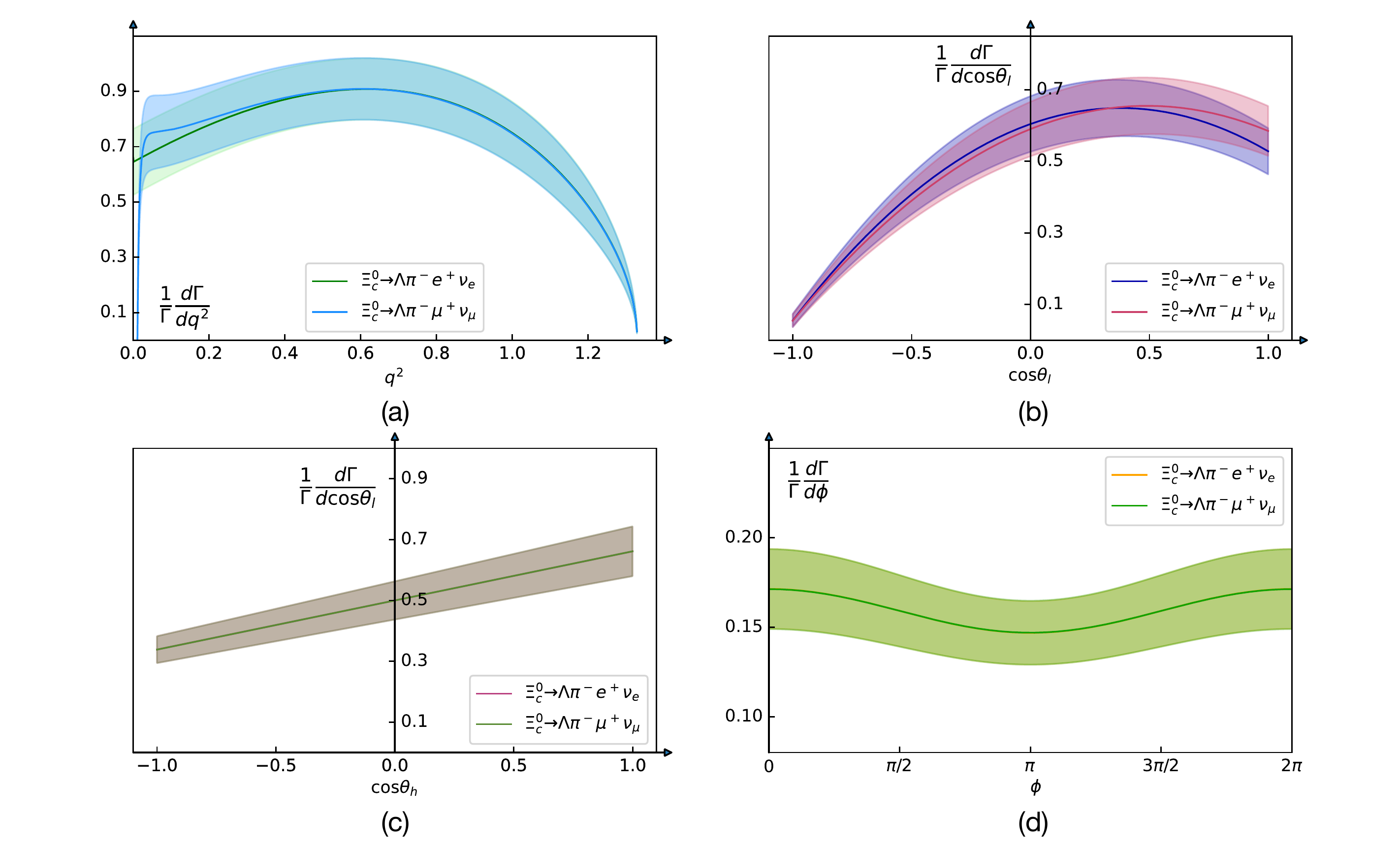}
%\caption{}
\centering
\caption{ Differential decay width  for $\Xi_{c}^0\rightarrow  \Lambda \pi^{-}\ell^{+}\nu_{\ell}$  as function of $q^{2}$ (a), $\cos\theta_{l}$ (b),  $\cos\theta_{h}$ (c)  and $\phi$ (d). Note that in sub-fig.(b) and (c), the errors increasing with $\theta_{l}$ form -1 to 1 and $\theta_{h}$ from 1 to -1, as a results of the remarkable uncertainties of form factors at small $q^2$ region. }
\label{Xi angular distributionA}
\end{figure}

 \item $\Omega_{c}^0\rightarrow\Lambda K^-\ell^+\nu_{\ell}$ 
 
 Due to the absence of lattice QCD calculation,  we use $\Omega_{c}^0\rightarrow\Omega^{-}$ transition form factors calculated by light-front quark model \cite{Hsiao:2020gtc}. The form factors can be expressed as the following double-pole form:
\begin{align}
F(q^{2})=\frac{F(0)}{1-a(q^2/m^{2}_{F})+b(q^4/m^{4}_{F})}\;,
\end{align}
where F(0) is the value of the form factors at $q^2=0$ with $m_{F}=1.86$ GeV, $\delta\equiv\delta{m_c}/m_c=\pm0.04$. The numerical results from light-front quark model are collected in Table~\ref{Omegac form factor}.

   \begin{table}[!htbp]
  \centering
  \begin{tabular}{c|c|c|c|c|c|c|c|c}
  \hline
  \hline
&$f_{1}$&$f_{2}$&$f_{3}$&$f_{4}$&$g_{1}$&$g_{2}$&$g_{3}$&$g_{4}$ \\ \hline
$F(0)$&$0.54+0.13\delta$&$0.35-0.36\delta$&$0.33+0.59\delta$&$0.97+0.22\delta$&$2.05+1.38\delta$&$-0.06+0.33\delta$&$-1.32-0.32\delta$&$-0.44+0.11\delta$  \\ \hline
a&-0.27&-30.0&0.96&-0.53&-3.66&-1.15&-4.01&-1.29 \\ \hline
b&1.65&96.82&9.25&1.41&1.41&71.66&5.68&-0.58 \\ \hline
\hline
  \end{tabular}
  \caption{$\Omega^{0}_{c}\rightarrow\Omega^{-}$ transition form factors with F(0) at $q^{2}=0$.}~\label{Omegac form factor}
\end{table}

 Through the decay width of $\Omega_{c}$ in Eq.(\ref{eq:omegafourbody}) and integrating out all variables, we can obtain the  numerical results  of decay widths and branching fractions  with errors:
  \begin{align}
 & \Omega_{c}\rightarrow \Lambda K^{-}e^{+}\nu_{e}:~~  \Gamma=8.889(344)\times 10^{-15} s^{-1} , ~~  \mathcal{B}=0.362(14)\%,   \\
 & \Omega_{c}\rightarrow \Lambda K^{-}\mu^{+}\nu_{\nu}:~~ \Gamma=8.771(343)\times 10^{-15} s^{-1}, ~~  \mathcal{B}=0.350(14)\%.
  \end{align}

Fig.~\ref{Xi angular distributionA}(a) shows  the differential decay widths of $\Omega_{c}^0\rightarrow \Lambda K^{-}\ell^{+}\nu_{\ell}$ as function of $q^{2}$. We also give the results of  angular distributions of total decay width, the numerical results are listed in Eq.(\ref{coee},\ref{coemu}), and in Fig.\ref{Xi angular distributionA} the illustration of the angular distributions with $\phi$, $\cos\theta_{l}$ and $\cos\theta_{h}$ are displayed.

 \begin{enumerate}
\item $\Omega_{c}^0\rightarrow \Lambda K^{-} e^{+}\nu_{e}$ 
\begin{align}
\frac{d\Gamma}{\Gamma d\cos\theta_{l}}=&~0.472(2)+0.244(3)\cos\theta_{l}-0.085(14)\cos2\theta_{l} \; ,\nonumber \\
\frac{d\Gamma}{\Gamma d\cos\theta_{h}}=&~0.548(24)-0.002(0)\cos\theta_{h}+0.143(13)\cos2\theta_{h}-0.001(0)\cos3\theta_{h}\; ,\nonumber \\
\frac{d\Gamma}{\Gamma d\phi}=&~0.159(6)-0.0233(8)\cos\theta_{\phi}-0.0173(4)\cos2\theta_{\phi}\; .
\label{coee}
\end{align}

\item $\Omega_{c}^0\rightarrow \Lambda K^{-} \mu^{+}\nu_{\mu}$ 
\begin{align}
\frac{d\Gamma}{\Gamma d\cos\theta_{l}}=&~0.479(16)+0.273(5)\cos\theta_{l}-0.064(11)\cos2\theta_{l} \; ,\nonumber \\
\frac{d\Gamma}{\Gamma d\cos\theta_{h}}=&~0.5479(238)-0.0021(6)\cos\theta_{h}+0.1436(131)\cos2\theta_{h}-0.0014(0)\cos3\theta_{h}\; ,\nonumber \\
\frac{d\Gamma}{\Gamma d\phi}=&~0.1596(62)-0.0224(8)\cos\theta_{\phi}-0.0120(2)\cos2\theta_{\phi}\; .
\label{coemu}
\end{align}
 \end{enumerate}

\begin{figure}[htbp]
\centering
\includegraphics[width=1.0\columnwidth]{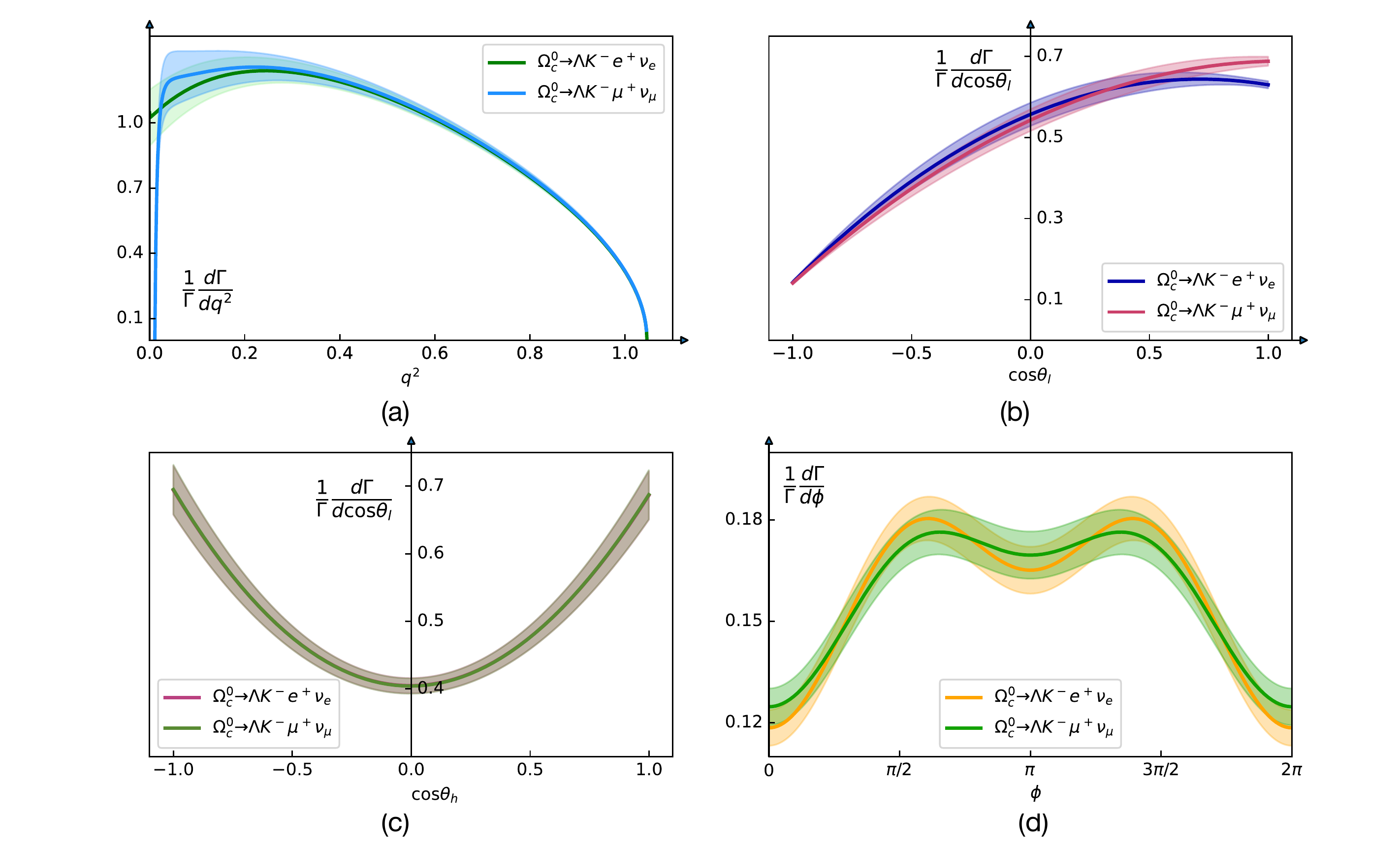}
%\caption{}
\centering
\caption{ Dfferential decay width  for $\Omega_{c}\rightarrow  \Lambda K^{-}\ell^{+}\nu_{\ell}$  as function of $q^{2}$ (a),  $\cos\theta_{l}$ (b),  $\cos\theta_{h}$ (c) and $\phi$ (d). Note that in sub-fig.(b) and (c), the errors increasing with $\theta_{l}$ form -1 to 1 and $\theta_{h}$ from 1 to -1, as a results of the remarkable uncertainties of form factors at small $q^2$ region. }
\label{Xi angular distributionA}
\end{figure}

\end{itemize}

\section{Summary}

In summary,  we have studied charmed baryon four-body semileptonic decays  including the antitriplet charmed baryons ($\Lambda_{c}^{+}$, $\Xi^{0}_{c}$) and sextet $\Omega^{0}_{c}$. With the form factors of the $\Lambda^{+}_{c}\rightarrow\Lambda$ and $\Xi^{0}_{c}\rightarrow\Xi^{-}$ calculated by lattice QCD and   the light-front quark model calculation of  $\Omega^{0}_{c}\rightarrow\Omega^{-}$ form factors, we have predicted $\Gamma(\Lambda_{c}^{+} \rightarrow p \pi^{-} e^{+} \nu_{e})=8.061(482)\times 10^{-14} s^{-1}$, $\Gamma(\Lambda_{c}^+\rightarrow p \pi^{-}\mu^{+}\nu_{\mu})=8.126(462)\times 10^{-14} s^{-1}$, $\Gamma(\Xi_{c}\rightarrow \Lambda \pi^{-}e^{+}\nu_{e})=1.031(129)\times 10^{-13} s^{-1}$, $\Gamma(\Xi_{c}\rightarrow \Lambda \pi^{-}\mu^{+}\nu_{\nu})=1.037(131)\times 10^{-13} s^{-1}$, $\Gamma(\Omega_{c}\rightarrow \Lambda K^{-}e^{+}\nu_{e})=8.889(344)\times 10^{-15} s^{-1}$, $\Gamma(\Omega_{c}\rightarrow \Lambda K^{-}\mu^{+}\nu_{\nu})=8.771(343)\times 10^{-15} s^{-1}$, and $\mathcal{B}(\Lambda_{c}^{+} \rightarrow p \pi^{-} e^{+} \nu_{e})=2.48(15)\%$, $\mathcal{B}(\Lambda_{c}^+\rightarrow p \pi^{-}\mu^{+}\nu_{\mu})=2.50(14)\%$,  $\mathcal{B}(\Xi_{c}\rightarrow \Lambda \pi^{-}e^{+}\nu_{e})=2.40(30)\%$, $\mathcal{B}(\Xi_{c}\rightarrow \Lambda \pi^{-}\mu^{+}\nu_{\nu})=2.41(30)\%$, $\mathcal{B}(\Omega_{c}\rightarrow \Lambda K^{-}e^{+}\nu_{e})=0.362(14)\%$, $\mathcal{B}(\Omega_{c}\rightarrow \Lambda K^{-}\mu^{+}\nu_{\nu})=0.350(14)\%$, using helicity amplitude technique. In addition, we give  the angular distributions of all the process with different angulars $\cos\theta_{l}$,$\cos\theta_{h}$ and $\phi$. In the future, we expect to study the more general cases of semileptonic charmed baryon decays by calculating the form factors such as $\Xi_c\to\Lambda$ and so on. This work can provide a   theoretical  basis for the ongoing experiments at BESIII, LHCb and BELLE-II.

\section*{ACKNOWLEDGMENT}
We greatly thank Prof. Wei Wang for inspiration and valuable discussions, and thank Ji Xu and Zhen-Xing Zhao for valuable discussions. This work is supported by  National Natural Science Foundation of China under the Grant No. 11735010, 12005130, U2032102,  and the China Postdoctoral Science Foundation and the National Postdoctoral Program for Innovative Talents (Grant No. BX20190207).

\section*{APPENDIX}

\subsection{Definitions of spinors, polarization vectors and vectorial spinors}\label{spinors}
In this part, we will introduce the spinors and polarization vectors of fermions and vector boson~\cite{Auvil:1966eao}. The spinors for spin-1/2 fermions are
\begin{align}
u\left(\vec{p},+\frac{1}{2}\right)=&N_{p}\left(\begin{array}{c}
 e^{-i \phi / 2} \cos \left(\theta / 2\right) \\
 e^{i \phi / 2} \sin \left(\theta / 2\right) \\
\frac{\left| \vec{p}\right|}{E+m} e^{-i \phi / 2} \cos \left(\theta / 2\right) \\
\frac{\left| \vec{p}\right|}{E+m} e^{i \phi / 2} \sin \left(\theta / 2\right)
\end{array}\right), \\
u\left(\vec{p},-\frac{1}{2}\right)=&N_{p}\left(\begin{array}{c}
 -e^{-i \phi / 2} \sin \left(\theta / 2\right) \\
 e^{i \phi / 2} \cos \left(\theta / 2\right) \\
\frac{\left| \vec{p}\right|}{E+m} e^{-i \phi / 2} \sin \left(\theta / 2\right) \\
-\frac{\left| \vec{p}\right|}{E+m} e^{i \phi / 2} \cos \left(\theta / 2\right)
\end{array}\right), \\
\end{align}
where $N_{p}=\sqrt{E+m}$ and $p^{\mu}=(p^{0},\vec{p}\sin \theta \cos \phi,\vec{p}\sin \theta\sin \phi,\vec{p}\cos\theta)$. The spinors for anti-fermions can be obtained by $\bar{v}(\vec{p},s)=-i \gamma^2 u^{\star}(\vec{p},s)$. The polarization vectors can be expressed as
\begin{align}
	\epsilon^{\mu}(+1)=&\frac{1}{\sqrt{2}}(0,-\cos \theta \cos \phi + i \sin \phi,-\cos \theta \sin \phi -i\cos \phi, \sin \theta)\; , \nonumber \\
	\epsilon^{\mu}(-1)=&\frac{1}{\sqrt{2}}(0,\cos \theta \cos \phi + i \sin \phi,\cos \theta \sin \phi -i\cos \phi, -\sin \theta)\; , \nonumber \\
	\epsilon^{\mu}(0)=&\frac{1}{\sqrt{m}}(\left|\vec{p}\right|,p^{0}\sin \theta \cos \phi , p^{0}\sin \theta \sin \phi , p^{0}\cos \theta)\; .
\end{align}

 The vectorial spinor $U_{\alpha}(\vec{p},S_z)$ for spin-$3/2$ baryon is given as
 \begin{align}
 	U_{\alpha}\left(\vec{p}, S_{z}\right)=\sum_{s_{z}, s_{z}^{\prime}}\left\langle s=1, s_{z} ; s^{\prime}=\frac{1}{2}, s_{z}^{\prime} \mid s=1, s^{\prime}=\frac{1}{2} ; S=\frac{3}{2}, S_{z}\right\rangle \varepsilon_{\alpha}\left(\vec{p}, s_{z}\right) u\left(\vec{p}, s_{z}^{\prime}\right),
 \end{align}
 where $u(\vec{p},s_z')$ is wave function for spin-$1/2$ and $\epsilon_{\alpha}(\vec{p},s_z)$ for spin-1. And the Clebsch-Gordan coefficients
 \begin{align}
S_{z}=\frac{3}{2}:~~~ &\left\langle s_{z}=1, s_{z}^{\prime}=\frac{1}{2} \mid S_{z}=\frac{3}{2}\right\rangle=1, \\
S_{z}=\frac{1}{2}:~~~ &\left\langle s_{z}=1, s_{z}^{\prime}=-\frac{1}{2} \mid S_{z}=\frac{1}{2}\right\rangle=\frac{1}{\sqrt{3}}, \quad\left\langle s_{z}=0, s_{z}^{\prime}=\frac{1}{2} \mid S_{z}=\frac{1}{2}\right\rangle=\sqrt{\frac{2}{3}}, \\
S_{z}=-\frac{1}{2} :~~~ & \left\langle s_{z}=0, s_{z}^{\prime}=-\frac{1}{2} \mid S_{z}=-\frac{1}{2}\right\rangle=\sqrt{\frac{2}{3}}, \quad\left\langle s_{z}=-1, s_{z}^{\prime}=\frac{1}{2} \mid S_{z}=-\frac{1}{2}\right\rangle=\frac{1}{\sqrt{3}}, \\
S_{z}=-\frac{3}{2}:~~~ & \left\langle s_{z}=-1, s_{z}^{\prime}=-\frac{1}{2} \mid S_{z}=-\frac{3}{2}\right\rangle=1.
\end{align}

%%%%

\subsection{Transformations of various-based form factors} \label{app:trans}

 Apart from the generalized parametrization in Eq.(\ref{spinhalfV},\ref{spinhalfA}) for spin-1/2-to-spin-1/2 case, another commonly treatment named helicity-based form factors,
 \begin{align}
\left\langle B_{2}\left(p_{2}, s_{2}\right)\left|V^{\mu}\right| B_{1}\left(p_{1}, s_{1}\right)\right\rangle=\bar{u}(&\left.p_{2}, s_{2}\right)\left[\left(m_{1}-m_{2}\right) \frac{q^{\mu}}{q^{2}} f_{0}\left(q^{2}\right)\right. \nonumber\\
&+\frac{m_{1}+m_{2}}{s_{+}}\left(p_{1}^{\mu}+p_{2}^{\mu}-\left(m_{1}^{2}-m_{2}^{2}\right) \frac{q^{\mu}}{q^{2}}\right) f_{+}\left(q^{2}\right) \nonumber\\
&\left.+\left(\gamma^{\mu}-\frac{2 m_{2}}{s_{+}} p_{1}^{\mu}-\frac{2 m_{1}}{s_{+}} p_{2}^{\mu}\right) f_{\perp}\left(q^{2}\right)\right] u\left(p_{1}, s_{1}\right), \\
\left\langle B_{2}\left(p_{2}, s_{2}\right)\left|A^{\mu}\right| B_{1}\left(p_{1}, s_{1}\right)\right\rangle=-\bar{u} &\left(p_{2}, s_{2}\right) \gamma_{5}\left[\left(m_{1}+m_{2}\right) \frac{q^{\mu}}{q^{2}} g_{0}\left(q^{2}\right)\right.\nonumber\\
&+\frac{m_{1}-m_{2}}{s_{-}}\left(p_{1}^{\mu}+p_{2}^{\mu}-\left(m_{1}^{2}-m_{2}^{2}\right) \frac{q^{\mu}}{q^{2}}\right) g_{+}\left(q^{2}\right) \nonumber\\
&\left.+\left(\gamma^{\mu}-\frac{2 m_{2}}{s_{-}} p_{1}^{\mu}-\frac{2 m_{1}}{s_{-}} p_{2}^{\mu}\right) g_{\perp}\left(q^{2}\right)\right] u\left(p_{1}, s_{1}\right),
\end{align}
one can easily obtain the helicity-based form factors from the generalized ones under the following transformation:
\begin{align}
\left(\begin{array}{c}
f_{\perp}\left(q^{2}\right) \\
f_{0}\left(q^{2}\right) \\
f_{+}\left(q^{2}\right)
\end{array}\right)=\left(\begin{array}{ccc}
1 & -\frac{m_{1}+m_{2}}{m_{1}} & 0 \\
1 & 0 & \frac{q^{2}}{m_{1}\left(m_{1}-m_{2}\right)} \\
1 & -\frac{q^{2}}{m_{1}\left(m_{1}+m_{2}\right)} & 0
\end{array}\right)\left(\begin{array}{c}
f_{1}\left(q^{2}\right) \\
f_{2}\left(q^{2}\right) \\
f_{3}\left(q^{2}\right)
\end{array}\right), \\
%%%%%%%%%%%%%
\left(\begin{array}{c}
g_{\perp}\left(q^{2}\right) \\
g_{0}\left(q^{2}\right) \\
g_{+}\left(q^{2}\right)
\end{array}\right)=\left(\begin{array}{ccc}
1 & \frac{m_{1}-m_{2}}{m_{1}} & 0 \\
1 & 0 & -\frac{q^{2}}{m_{1}\left(m_{1}+m_{2}\right)} \\
1 & \frac{q^{2}}{m_{1}\left(m_{1}-m_{2}\right)} & 0
\end{array}\right)\left(\begin{array}{l}
g_{1}\left(q^{2}\right) \\
g_{2}\left(q^{2}\right) \\
g_{3}\left(q^{2}\right)
\end{array}\right).
\end{align}

The helicity-based parametrized form factors of $\Omega_c^0\to\Omega^-$ are given as
\begin{align}
 \langle B^{\prime}_{2}(p_2,s_2)|\bar{s}\gamma^{\mu}c|B^{\prime}_{1}(p_1,s_1)\rangle =-\bar{u}(&p_2,s_2)\gamma_{5}\left[f_{0}\frac{m_{2}}{s_{-}}\frac{(m_{2}+m_{1})p_{1}^{\alpha}q^{\mu}}{q^2}\right. \nonumber \\
&+f_{+}\frac{m_{1}}{s_{+}}\frac{(m_{1}-m_{2})p^{\alpha}\big(q^2(p_{1}^{\mu}+p_{2}^{\mu})-(m_{2}^{2}-m_{1}^{2})q^{\mu}\big)}{q^2 s_{+}}\nonumber \\
&+f_{\perp}\frac{m_{1}}{s_{+}}\left(p^{\alpha}\gamma^{\mu}-\frac{2p^{\alpha}(m_{2}p_{2}^{\mu}-m_{1}p_{1}^{\mu})}{s_{+}}\right) \nonumber \\
&+\left.f_{\perp^{'}}\frac{m_{1}}{s_{+}}\left(p^{\alpha}\gamma^{\mu}+\frac{2p_{1}^{\alpha}p_{2}^{\mu}}{m_{1}}+\frac{2p_{1}^{\alpha}(m_{1}p_{2}^{\mu}-m_{2}p^{\mu})}{s_{+}}\right)-\frac{s_{+}g^{\alpha\mu}}{m_{2}}\right]u(p_1,s_1)\; ,\nonumber \\
%%%
\langle B_{2}^{\prime}(p_2,s_2)|\bar{s}\gamma^{\mu}\gamma_{5}c|B_{1}^{\prime}(p_1,s_1)\rangle=\bar{u}(&p_2,s_2)\left[g_{0}\frac{m_{2}}{s_{+}}\frac{(m_{2}-m_{1})p_{1}^{\alpha}q^{\mu}}{q^2}\right. \nonumber \\
&+g_{+}\frac{m_{2}}{s_{-}}\frac{(m_{2}+m_{1})p^{\alpha}\big(q^2(p_{1}^{\mu}+p_{2}^{\mu})-(m_{1}^{2}-m_{2}^{2})q^{\mu}\big)}{q^2 s_{-}}\nonumber \\
&+g_{\perp}\frac{m_{2}}{s_{-}}\left(p^{\alpha}\gamma^{\mu}-\frac{2p^{\alpha}(m_{1}p_{2}^{\mu}+m_{2}p_{1}^{\mu})}{s_{+}}\right)\nonumber \\
&+\left.g_{\perp^{'}}\frac{m_{2}}{s_{-}}\left(p^{\alpha}\gamma^{\mu}-\frac{2p_{1}^{\alpha}p_{2}^{\mu}}{m_{2}}+\frac{2p_{1}^{\alpha}(m_{1}p_{2}^{\mu}+m_{2}p^{\mu})}{s_{-}}\right)+\frac{s_{-}g^{\alpha\mu}}{m_{2}}\right]u(p_1,s_1)\; .\nonumber \\
\end{align}
Similarly, we can obtain the helicity-based form factors from the generalized ones in Eq.(\ref{spinonehalfV},\ref{spinonehalfA}) under the following transformation
\begin{align}
\left(\begin{array}{c}
f_{\perp}\left(q^{2}\right) \\
f^{\prime}_{\perp}\left(q^{2}\right) \\
f_{0}\left(q^{2}\right) \\
f_{+}\left(q^{2}\right)
\end{array}\right)=\left(\begin{array}{cccc}
\frac{s_{+}}{m_{1}m_{2}} &0 & 0 &-1 \\
0 & 0 & 0 &1 \\
\frac{\left(m_{1}-m_{2}\right)^{2}-q^{2}}{m_{1}m_{2}} & -\frac{\left(m_{1}^{2}-m_{2}^{2}+q^{2}\right)s_{-}}{2m^{2}_{1}m_{2}(m_{1}+m_{2})} & \frac{\left(-m_{1}^{2}+m_{2}^{2}+q^{2}\right)s_{-}}{2m_{1}m^{2}_{2}(m_{1}+m_{2})} & \frac{-2\left(m_{1}-m_{2}\right)^{2}+2q^{2}+\frac{(-m^{2}_{1}+m^{2}_{2}+q^{2})s_{-}}{m_{2}(m_{1}+m_{2})}}{s_{+}} \\
\frac{s_{+}}{m_{1}m_{2}} & -\frac{s_{-}s_{+}}{2m^{3}_{1}m_{2}-2m^{2}_{1}m^{2}_{2}} & -\frac{s_{-}s_{+}}{2m^{2}_{1}m^{2}_{2}-2m_{1}m^{3}_{2}} &-2-\frac{s_{-}}{\left(m_{1}-m_{2}\right)m_{2}}
\end{array}\right)\left(\begin{array}{c}
f_{1}\left(q^{2}\right) \\
f_{2}\left(q^{2}\right) \\
f_{3}\left(q^{2}\right) \\
f_{4}\left(q^{2}\right)
\end{array}\right), \\
%%%%%%%%%%%%%
\left(\begin{array}{c}
g_{\perp}\left(q^{2}\right) \\
g^{\prime}_{\perp}\left(q^{2}\right) \\
g_{0}\left(q^{2}\right) \\
g_{+}\left(q^{2}\right) 
\end{array}\right)=\left(\begin{array}{cccc}
\frac{s_{-}}{m_{1}m_{2}} &0 & 0 &-1 \\
0 & 0 & 0 &1 \\
\frac{\left(m_{1}+m_{2}\right)^{2}-q^{2}}{m_{1}m_{2}} & \frac{\left(m_{1}^{2}-m_{2}^{2}+q^{2}\right)s_{+}}{2m^{2}_{1}m_{2}(m_{1}-m_{2})} & \frac{\left(m_{1}^{2}-m_{2}^{2}-q^{2}\right)s_{-}}{2m_{1}m^{2}_{2}(m_{1}-m_{2})} & \frac{-2\left(m_{1}+m_{2}\right)^{2}+2q^{2}+\frac{(m^{2}_{1}-m^{2}_{2}-q^{2})s_{+}}{m_{2}(m_{1}-m_{2})}}{s_{-}} \\
\frac{s_{-}}{m_{1}m_{2}} & \frac{s_{-}s_{+}}{2m^{3}_{1}m_{2}+2m^{2}_{1}m^{2}_{2}} & \frac{s_{-}s_{+}}{2m^{2}_{1}m^{2}_{2}+2m_{1}m^{3}_{2}} &-2+\frac{s_{+}}{\left(m_{1}+m_{2}\right)m_{2}}
\end{array}\right)\left(\begin{array}{l}
g_{1}\left(q^{2}\right) \\
g_{2}\left(q^{2}\right) \\
g_{3}\left(q^{2}\right) \\
g_{4}\left(q^{2}\right)
\end{array}\right).
\end{align}


\begin{thebibliography}{1}

%\cite{Wei:2009zv}
\bibitem{Wei:2009zv}
J.~T.~Wei \textit{et al.} [Belle],
%``Measurement of the Differential Branching Fraction and Forward-Backword Asymmetry for $B \to K^{(*)}\ell^+\ell^-$,''
Phys. Rev. Lett. \textbf{103}, 171801 (2009)
doi:10.1103/PhysRevLett.103.171801
[arXiv:0904.0770 [hep-ex]].
%522 citations counted in INSPIRE as of 26 Jun 2021

%\cite{Aaij:2015esa}
\bibitem{Aaij:2015esa}
R.~Aaij \textit{et al.} [LHCb],
%``Angular analysis and differential branching fraction of the decay $B^0_s\to\phi\mu^+\mu^-$,''
JHEP \textbf{09}, 179 (2015)
doi:10.1007/JHEP09(2015)179
[arXiv:1506.08777 [hep-ex]].
%415 citations counted in INSPIRE as of 26 Jun 2021

%\cite{Aaij:2015oid}
\bibitem{Aaij:2015oid}
R.~Aaij \textit{et al.} [LHCb],
%``Angular analysis of the $B^{0} \to K^{*0} \mu^{+} \mu^{-}$ decay using 3 fb$^{-1}$ of integrated luminosity,''
JHEP \textbf{02}, 104 (2016)
doi:10.1007/JHEP02(2016)104
[arXiv:1512.04442 [hep-ex]].
%726 citations counted in INSPIRE as of 29 Jun 2021

%\cite{Aaij:2017vbb}
\bibitem{Aaij:2017vbb}
R.~Aaij \textit{et al.} [LHCb],
%``Test of lepton universality with $B^{0} \rightarrow K^{*0}\ell^{+}\ell^{-}$ decays,''
JHEP \textbf{08}, 055 (2017)
doi:10.1007/JHEP08(2017)055
[arXiv:1705.05802 [hep-ex]].
%852 citations counted in INSPIRE as of 29 Jun 2021

%\cite{Aaij:2020nrf}
\bibitem{Aaij:2020nrf}
R.~Aaij \textit{et al.} [LHCb],
%``Measurement of $CP$-Averaged Observables in the $B^{0}\rightarrow K^{*0}\mu^{+}\mu^{-}$ Decay,''
Phys. Rev. Lett. \textbf{125}, no.1, 011802 (2020)
doi:10.1103/PhysRevLett.125.011802
[arXiv:2003.04831 [hep-ex]].
%91 citations counted in INSPIRE as of 26 Jun 2021

%\cite{CroninHennessy:2000bz}
\bibitem{CroninHennessy:2000bz} 
  D.~Cronin-Hennessy {\it et al.} [CLEO Collaboration],
  %``Observation of the Omega0(c) charmed baryon at CLEO,''
  Phys.\ Rev.\ Lett.\  {\bf 86}, 3730 (2001)
  doi:10.1103/PhysRevLett.86.3730
  [hep-ex/0010035].
  %%CITATION = doi:10.1103/PhysRevLett.86.3730;%%
  %25 citations counted in INSPIRE as of 25 May 2021
  
  %\cite{Ammar:2002pf}
\bibitem{Ammar:2002pf} 
  R.~Ammar {\it et al.} [CLEO Collaboration],
  %``Observation of the decay Omega0(c) ---> Omega- e+ nu(e),''
  Phys.\ Rev.\ Lett.\  {\bf 89}, 171803 (2002)
  doi:10.1103/PhysRevLett.89.171803
  [hep-ex/0207078].
  %%CITATION = doi:10.1103/PhysRevLett.89.171803;%%
  %14 citations counted in INSPIRE as of 25 May 2021
  
  %\cite{Aubert:2007bt}
\bibitem{Aubert:2007bt} 
  B.~Aubert {\it et al.} [BaBar Collaboration],
  %``Production and decay of Omega0(c),''
  Phys.\ Rev.\ Lett.\  {\bf 99}, 062001 (2007)
  doi:10.1103/PhysRevLett.99.062001
  [hep-ex/0703030 [HEP-EX]].
  %%CITATION = doi:10.1103/PhysRevLett.99.062001;%%
  %31 citations counted in INSPIRE as of 25 May 2021
  
  %\cite{Yelton:2017uzv}
\bibitem{Yelton:2017uzv} 
  J.~Yelton {\it et al.} [Belle Collaboration],
  %``Measurement of branching fractions of hadronic decays of the $\Omega_c^0$  baryon,''
  Phys.\ Rev.\ D {\bf 97}, no. 3, 032001 (2018)
  doi:10.1103/PhysRevD.97.032001
  [arXiv:1712.01333 [hep-ex]].
  %%CITATION = doi:10.1103/PhysRevD.97.032001;%%
  %9 citations counted in INSPIRE as of 25 May 2021
  
  %\cite{Tanabashi:2018oca}
\bibitem{Tanabashi:2018oca} 
  M.~Tanabashi {\it et al.} [Particle Data Group],
  %``Review of Particle Physics,''
  Phys.\ Rev.\ D {\bf 98}, no. 3, 030001 (2018).
  doi:10.1103/PhysRevD.98.030001
  %%CITATION = doi:10.1103/PhysRevD.98.030001;%%
  %6862 citations counted in INSPIRE as of 25 May 2021

%%%%%%charmed baryons

%\cite{Richman:1995wm}
\bibitem{Richman:1995wm} 
  J.~D.~Richman and P.~R.~Burchat,
  %``Leptonic and semileptonic decays of charm and bottom hadrons,''
  Rev.\ Mod.\ Phys.\  {\bf 67}, 893 (1995)
  doi:10.1103/RevModPhys.67.893
  [hep-ph/9508250].
  %%CITATION = doi:10.1103/RevModPhys.67.893;%%
  %282 citations counted in INSPIRE as of 25 May 2021
  
  %\cite{Eichten:1989zv}
\bibitem{Eichten:1989zv} 
  E.~Eichten and B.~R.~Hill,
  %``An Effective Field Theory for the Calculation of Matrix Elements Involving Heavy Quarks,''
  Phys.\ Lett.\ B {\bf 234}, 511 (1990).
  doi:10.1016/0370-2693(90)92049-O
  %%CITATION = doi:10.1016/0370-2693(90)92049-O;%%
  %1108 citations counted in INSPIRE as of 25 May 2021
  
  %\cite{Neubert:1993mb}
\bibitem{Neubert:1993mb} 
  M.~Neubert,
  %``Heavy quark symmetry,''
  Phys.\ Rept.\  {\bf 245}, 259 (1994)
  doi:10.1016/0370-1573(94)90091-4
  [hep-ph/9306320].
  %%CITATION = doi:10.1016/0370-1573(94)90091-4;%%
  %1637 citations counted in INSPIRE as of 25 May 2021
  
  %%%%%%%study in charmed baryons in theory
  
  
  %\cite{Cheng:1991sn}
\bibitem{Cheng:1991sn} 
  H.~Y.~Cheng and B.~Tseng,
  %``Nonleptonic weak decays of charmed baryons,''
  Phys.\ Rev.\ D {\bf 46}, 1042 (1992)
  Erratum: [Phys.\ Rev.\ D {\bf 55}, 1697 (1997)].
  doi:10.1103/PhysRevD.55.1697, 10.1103/PhysRevD.46.1042
  %%CITATION = doi:10.1103/PhysRevD.55.1697, 10.1103/PhysRevD.46.1042;%%
  %98 citations counted in INSPIRE as of 25 May 2021
  
  %\cite{Gronau:2013mza}
\bibitem{Gronau:2013mza} 
  M.~Gronau and J.~L.~Rosner,
  %``Flavor SU(3) and $\Lambda_b$ decays,''
  Phys.\ Rev.\ D {\bf 89}, no. 3, 037501 (2014)
  Erratum: [Phys.\ Rev.\ D {\bf 91}, no. 11, 119902 (2015)]
  doi:10.1103/physrevd.91.119902, 10.1103/PhysRevD.89.037501
  [arXiv:1312.5730 [hep-ph]].
  %%CITATION = doi:10.1103/physrevd.91.119902, 10.1103/PhysRevD.89.037501;%%
  %18 citations counted in INSPIRE as of 25 May 2021
  
  
    %%%%%%%study in charmed baryons in experiments
    
    
      %\cite{Ablikim:2015prg}
\bibitem{Ablikim:2015prg} 
  M.~Ablikim {\it et al.} [BESIII Collaboration],
  %``Measurement of the absolute branching fraction for $\Lambda^+_{c}\to \Lambda e^+\nu_e$,''
  Phys.\ Rev.\ Lett.\  {\bf 115}, no. 22, 221805 (2015)
  doi:10.1103/PhysRevLett.115.221805
  [arXiv:1510.02610 [hep-ex]].
  %%CITATION = doi:10.1103/PhysRevLett.115.221805;%%
  %66 citations counted in INSPIRE as of 25 May 2021
  
  %\cite{Ablikim:2016mcr}
\bibitem{Ablikim:2016mcr} 
  M.~Ablikim {\it et al.} [BESIII Collaboration],
  %``Observation of $\Lambda^+_{c}\to nK^0_S\pi^+$,''
  Phys.\ Rev.\ Lett.\  {\bf 118}, no. 11, 112001 (2017)
  doi:10.1103/PhysRevLett.118.112001
  [arXiv:1611.02797 [hep-ex]].
  %%CITATION = doi:10.1103/PhysRevLett.118.112001;%%
  %31 citations counted in INSPIRE as of 25 May 2021
  
  
  %\cite{Ablikim:2016vqd}
\bibitem{Ablikim:2016vqd} 
  M.~Ablikim {\it et al.} [BESIII Collaboration],
  %``Measurement of the absolute branching fraction for $\Lambda_c^+\rightarrow \Lambda \mu^+\nu_{\mu}$,''
  Phys.\ Lett.\ B {\bf 767}, 42 (2017)
  doi:10.1016/j.physletb.2017.01.047
  [arXiv:1611.04382 [hep-ex]].
  %%CITATION = doi:10.1016/j.physletb.2017.01.047;%%
  %38 citations counted in INSPIRE as of 25 May 2021
  

  


  
  %\cite{Ablikim:2017ors}
\bibitem{Ablikim:2017ors} 
  M.~Ablikim {\it et al.} [BESIII Collaboration],
  %``Evidence for the singly-Cabibbo-suppressed decay $\Lambda_{c}^{+} \to p\eta$ and search for $\Lambda_{c}^{+} \to p\pi^{0}$,''
  Phys.\ Rev.\ D {\bf 95}, no. 11, 111102 (2017)
  doi:10.1103/PhysRevD.95.111102
  [arXiv:1702.05279 [hep-ex]].
  %%CITATION = doi:10.1103/PhysRevD.95.111102;%%
  %29 citations counted in INSPIRE as of 25 May 2021
  
  
  %\cite{Ablikim:2017iqd}
\bibitem{Ablikim:2017iqd} 
  M.~Ablikim {\it et al.} [BESIII Collaboration],
  %``Observation of the decay $\Lambda_c^+\rightarrow \Sigma^- \pi^+\pi^+\pi^0$,''
  Phys.\ Lett.\ B {\bf 772}, 388 (2017)
  doi:10.1016/j.physletb.2017.06.065
  [arXiv:1705.11109 [hep-ex]].
  %%CITATION = doi:10.1016/j.physletb.2017.06.065;%%
  %22 citations counted in INSPIRE as of 25 May 2021
  
  %\cite{Aaij:2017pgy}
\bibitem{Aaij:2017pgy} 
  R.~Aaij {\it et al.} [LHCb Collaboration],
  %``Measurement of branching fractions of charmless four-body $\Lambda_b^0$ and $\Xi_b^0$ decays,''
  JHEP {\bf 1802}, 098 (2018)
  doi:10.1007/JHEP02(2018)098
  [arXiv:1711.05490 [hep-ex]].
  %%CITATION = doi:10.1007/JHEP02(2018)098;%%
  %17 citations counted in INSPIRE as of 25 May 2021
  
  %\cite{Aaij:2017svr}
\bibitem{Aaij:2017svr} 
  R.~Aaij {\it et al.} [LHCb Collaboration],
  %``Measurement of the shape of the $\Lambda_b^0\to\Lambda_c^+ \mu^- \overline{\nu}_{\mu}$ differential decay rate,''
  Phys.\ Rev.\ D {\bf 96}, no. 11, 112005 (2017)
  doi:10.1103/PhysRevD.96.112005
  [arXiv:1709.01920 [hep-ex]].
  %%CITATION = doi:10.1103/PhysRevD.96.112005;%%
  %42 citations counted in INSPIRE as of 25 May 2021
  
  %\cite{Aaij:2017nsd}
\bibitem{Aaij:2017nsd} 
  R.~Aaij {\it et al.} [LHCb Collaboration],
  %``Search for the rare decay $\Lambda_{c}^{+} \to p\mu^+\mu^-$,''
  Phys.\ Rev.\ D {\bf 97}, no. 9, 091101 (2018)
  doi:10.1103/PhysRevD.97.091101
  [arXiv:1712.07938 [hep-ex]].
  %%CITATION = doi:10.1103/PhysRevD.97.091101;%%
  %22 citations counted in INSPIRE as of 25 May 2021
  
  %\cite{Aaij:2017xva}
\bibitem{Aaij:2017xva} 
  R.~Aaij {\it et al.} [LHCb Collaboration],
  %``A measurement of the $CP$ asymmetry difference in $\varLambda_{c}^{+} \to pK^{-}K^{+}$ and $p\pi^{-}\pi^{+}$ decays,''
  JHEP {\bf 1803}, 182 (2018)
  doi:10.1007/JHEP03(2018)182
  [arXiv:1712.07051 [hep-ex]].
  %%CITATION = doi:10.1007/JHEP03(2018)182;%%
  %23 citations counted in INSPIRE as of 25 May 2021
  
  
  %\cite{Aaij:2017rin}
\bibitem{Aaij:2017rin} 
  R.~Aaij {\it et al.} [LHCb Collaboration],
  %``Measurements of the branching fractions of $\Lambda_{c}^{+} \rightarrow p \pi^{-} \pi^{+}$, $\Lambda_{c}^{+} \rightarrow p K^{-} K^{+}$, and $\Lambda_{c}^{+} \rightarrow p \pi^{-} K^{+}$,''
  JHEP {\bf 1803}, 043 (2018)
  doi:10.1007/JHEP03(2018)043
  [arXiv:1711.01157 [hep-ex]].
  %%CITATION = doi:10.1007/JHEP03(2018)043;%%
  %15 citations counted in INSPIRE as of 25 May 2021
  
  
  %%%%%QCD sumrules%%%%%%%%%%%
  %\cite{Buras:1976dg}
\bibitem{Buras:1976dg} 
  A.~J.~Buras,
  %``Semileptonic Decays of Charmed Baryons,''
  Nucl.\ Phys.\ B {\bf 109}, 373 (1976).
  doi:10.1016/0550-3213(76)90241-8
  %%CITATION = doi:10.1016/0550-3213(76)90241-8;%%
  %46 citations counted in INSPIRE as of 25 May 2021
  
  %\cite{Gavela:1979wk}
\bibitem{Gavela:1979wk} 
  M.~B.~Gavela,
  %``Semileptonic Decays of Charmed Particles,''
  Phys.\ Lett.\  {\bf 83B}, 367 (1979).
  doi:10.1016/0370-2693(79)91129-8
  %%CITATION = doi:10.1016/0370-2693(79)91129-8;%%
  %29 citations counted in INSPIRE as of 25 May 2021
  
  %\cite{AvilaAoki:1989yi}
\bibitem{AvilaAoki:1989yi} 
  M.~Avila-Aoki, A.~Garcia, R.~Huerta and R.~Perez-Marcial,
  %``Predictions for Semileptonic Decays of Charm Baryons. 1. SU(4) Symmetry Limit,''
  Phys.\ Rev.\ D {\bf 40}, 2944 (1989).
  doi:10.1103/PhysRevD.40.2944
  %%CITATION = doi:10.1103/PhysRevD.40.2944;%%
  %17 citations counted in INSPIRE as of 25 May 2021
  
  
  %\cite{PerezMarcial:1989yh}
\bibitem{PerezMarcial:1989yh} 
  R.~Perez-Marcial, R.~Huerta, A.~Garcia and M.~Avila-Aoki,
  %``Predictions for Semileptonic Decays of Charm Baryons. 2. Nonrelativistic and {MIT} Bag Quark Models,''
  Phys.\ Rev.\ D {\bf 40}, 2955 (1989)
  Erratum: [Phys.\ Rev.\ D {\bf 44}, 2203 (1991)].
  doi:10.1103/PhysRevD.44.2203, 10.1103/PhysRevD.40.2955
  %%CITATION = doi:10.1103/PhysRevD.44.2203, 10.1103/PhysRevD.40.2955;%%
  %77 citations counted in INSPIRE as of 25 May 2021
  
  %\cite{Hussain:1990ai}
\bibitem{Hussain:1990ai} 
  F.~Hussain and J.~G.~Korner,
  %``Semileptonic charm baryon decays in the relativistic spectator quark model,''
  Z.\ Phys.\ C {\bf 51}, 607 (1991).
  doi:10.1007/BF01565586
  %%CITATION = doi:10.1007/BF01565586;%%
  %38 citations counted in INSPIRE as of 25 May 2021
  
  %\cite{Singleton:1990ye}
\bibitem{Singleton:1990ye} 
  R.~L.~Singleton,
  %``Semileptonic baryon decays with a heavy quark,''
  Phys.\ Rev.\ D {\bf 43}, 2939 (1991).
  doi:10.1103/PhysRevD.43.2939
  %%CITATION = doi:10.1103/PhysRevD.43.2939;%%
  %78 citations counted in INSPIRE as of 25 May 2021
  
  %\cite{Efimov:1991ex}
\bibitem{Efimov:1991ex} 
  G.~V.~Efimov, M.~A.~Ivanov and V.~E.~Lyubovitskij,
  %``Predictions for semileptonic decay rates of charmed baryons in the quark confinement model,''
  Z.\ Phys.\ C {\bf 52}, 149 (1991).
  doi:10.1007/BF01412338
  %%CITATION = doi:10.1007/BF01412338;%%
  %15 citations counted in INSPIRE as of 25 May 2021
  
  %\cite{Garcia:1992qe}
\bibitem{Garcia:1992qe} 
  A.~Garcia and R.~Huerta,
  %``Semileptonic decays of charmed baryons,''
  Phys.\ Rev.\ D {\bf 45}, 3266 (1992).
  doi:10.1103/PhysRevD.45.3266
  %%CITATION = doi:10.1103/PhysRevD.45.3266;%%
  %10 citations counted in INSPIRE as of 25 May 2021
  
  %\cite{Cheng:1995fe}
\bibitem{Cheng:1995fe} 
  H.~Y.~Cheng and B.~Tseng,
  %``1/M corrections to baryonic form-factors in the quark model,''
  Phys.\ Rev.\ D {\bf 53}, 1457 (1996)
  Erratum: [Phys.\ Rev.\ D {\bf 55}, 1697 (1997)]
  doi:10.1103/PhysRevD.53.1457, 10.1103/PhysRevD.55.1697.2
  [hep-ph/9502391].
  %%CITATION = doi:10.1103/PhysRevD.53.1457, 10.1103/PhysRevD.55.1697.2;%%
  %110 citations counted in INSPIRE as of 25 May 2021
  
  
  %\cite{Ivanov:1996fj}
\bibitem{Ivanov:1996fj} 
  M.~A.~Ivanov, V.~E.~Lyubovitskij, J.~G.~Korner and P.~Kroll,
  %``Heavy baryon transitions in a relativistic three quark model,''
  Phys.\ Rev.\ D {\bf 56}, 348 (1997)
  doi:10.1103/PhysRevD.56.348
  [hep-ph/9612463].
  %%CITATION = doi:10.1103/PhysRevD.56.348;%%
  %156 citations counted in INSPIRE as of 25 May 2021
  
  %\cite{Dosch:1997zx}
\bibitem{Dosch:1997zx} 
  H.~G.~Dosch, E.~Ferreira, M.~Nielsen and R.~Rosenfeld,
  %``Evidence from QCD sum rules for large violation of heavy quark symmetry in Lambda(b) semileptonic decay,''
  Phys.\ Lett.\ B {\bf 431}, 173 (1998)
  doi:10.1016/S0370-2693(98)00566-8
  [hep-ph/9712350].
  %%CITATION = doi:10.1016/S0370-2693(98)00566-8;%%
  %27 citations counted in INSPIRE as of 25 May 2021
  
  %\cite{Pervin:2005ve}
\bibitem{Pervin:2005ve} 
  M.~Pervin, W.~Roberts and S.~Capstick,
  %``Semileptonic decays of heavy lambda baryons in a quark model,''
  Phys.\ Rev.\ C {\bf 72}, 035201 (2005)
  doi:10.1103/PhysRevC.72.035201
  [nucl-th/0503030].
  %%CITATION = doi:10.1103/PhysRevC.72.035201;%%
  %69 citations counted in INSPIRE as of 25 May 2021
  
  
  %\cite{Liu:2009sn}
\bibitem{Liu:2009sn} 
  Y.~L.~Liu, M.~Q.~Huang and D.~W.~Wang,
  %``Improved analysis on the semi-leptonic decay Lambda(c) ---> Lambda l+ nu from QCD light-cone sum rules,''
  Phys.\ Rev.\ D {\bf 80}, 074011 (2009)
  doi:10.1103/PhysRevD.80.074011
  [arXiv:0910.1160 [hep-ph]].
  %%CITATION = doi:10.1103/PhysRevD.80.074011;%%
  %24 citations counted in INSPIRE as of 25 May 2021
  
  %\cite{Gutsche:2015rrt}
\bibitem{Gutsche:2015rrt} 
  T.~Gutsche, M.~A.~Ivanov, J.~G.~Korner, V.~E.~Lyubovitskij and P.~Santorelli,
  %``Semileptonic decays $\Lambda_c^+ \to \Lambda \ell^+ \nu_\ell\,\,(\ell=e,\mu)$ in the covariant quark model and comparison with the new absolute branching fraction measurements of Belle and BESIII,''
  Phys.\ Rev.\ D {\bf 93}, no. 3, 034008 (2016)
  doi:10.1103/PhysRevD.93.034008
  [arXiv:1512.02168 [hep-ph]].
  %%CITATION = doi:10.1103/PhysRevD.93.034008;%%
  %33 citations counted in INSPIRE as of 25 May 2021
  
  %\cite{Faustov:2016yza}
\bibitem{Faustov:2016yza} 
  R.~N.~Faustov and V.~O.~Galkin,
  %``Semileptonic decays of $\Lambda _c$ baryons in the relativistic quark model,''
  Eur.\ Phys.\ J.\ C {\bf 76}, no. 11, 628 (2016)
  doi:10.1140/epjc/s10052-016-4492-z
  [arXiv:1610.00957 [hep-ph]].
  %%CITATION = doi:10.1140/epjc/s10052-016-4492-z;%%
  %26 citations counted in INSPIRE as of 25 May 2021
  
  %%%%%lambdac%%%%
  

  
  
  %%%%%xic
  
  %\cite{Li:2018qak}
\bibitem{Li:2018qak} 
  Y.~B.~Li {\it et al.} [Belle Collaboration],
  %``First Measurements of Absolute Branching Fractions of the $\Xi_c^0$ Baryon at Belle,''
  Phys.\ Rev.\ Lett.\  {\bf 122}, no. 8, 082001 (2019)
  doi:10.1103/PhysRevLett.122.082001
  [arXiv:1811.09738 [hep-ex]].
  %%CITATION = doi:10.1103/PhysRevLett.122.082001;%%
  %33 citations counted in INSPIRE as of 25 May 2021
  
  %\cite{Zyla:2020zbs}
\bibitem{Zyla:2020zbs} 
  P.~A.~Zyla {\it et al.} [Particle Data Group],
  %``Review of Particle Physics,''
  PTEP {\bf 2020}, no. 8, 083C01 (2020).
  doi:10.1093/ptep/ptaa104
  %%CITATION = doi:10.1093/ptep/ptaa104;%%
  %1429 citations counted in INSPIRE as of 25 May 2021
  
  
  %%%%%%%%%%%xic experiment 2021%%%%%%%%%%%
  \bibitem{Y.B.Li:2021}
  Y.B. Li, C. P.Shen, et al.[Belle Collaboration],arXiv:2103.06496 [hep-ex]
  
  
  %%%%%%theory xic
  %\cite{Zhao:2018zcb}
\bibitem{Zhao:2018zcb} 
  Z.~X.~Zhao,
  %``Weak decays of heavy baryons in the light-front approach,''
  Chin.\ Phys.\ C {\bf 42}, no. 9, 093101 (2018)
  doi:10.1088/1674-1137/42/9/093101
  [arXiv:1803.02292 [hep-ph]].
  %%CITATION = doi:10.1088/1674-1137/42/9/093101;%%
  %36 citations counted in INSPIRE as of 25 May 2021
  
  %\cite{Azizi:2011mw}
\bibitem{Azizi:2011mw} 
  K.~Azizi, Y.~Sarac and H.~Sundu,
  %``Light cone QCD sum rules study of the semileptonic heavy $\Xi_{Q}$ and $\Xi'_{Q}$ transitions to $\Xi$ and $\Sigma $ baryons,''
  Eur.\ Phys.\ J.\ A {\bf 48}, 2 (2012)
  doi:10.1140/epja/i2012-12002-1
  [arXiv:1107.5925 [hep-ph]].
  %%CITATION = doi:10.1140/epja/i2012-12002-1;%%
  %21 citations counted in INSPIRE as of 25 May 2021
  
  %\cite{Geng:2018plk}
\bibitem{Geng:2018plk} 
  C.~Q.~Geng, Y.~K.~Hsiao, C.~W.~Liu and T.~H.~Tsai,
  %``Antitriplet charmed baryon decays with SU(3) flavor symmetry,''
  Phys.\ Rev.\ D {\bf 97}, no. 7, 073006 (2018)
  doi:10.1103/PhysRevD.97.073006
  [arXiv:1801.03276 [hep-ph]].
  %%CITATION = doi:10.1103/PhysRevD.97.073006;%%
  %44 citations counted in INSPIRE as of 25 May 2021
  
  %\cite{Geng:2019bfz}
\bibitem{Geng:2019bfz} 
  C.~Q.~Geng, C.~W.~Liu, T.~H.~Tsai and S.~W.~Yeh,
  %``Semileptonic decays of anti-triplet charmed baryons,''
  Phys.\ Lett.\ B {\bf 792}, 214 (2019)
  doi:10.1016/j.physletb.2019.03.056
  [arXiv:1901.05610 [hep-ph]].
  %%CITATION = doi:10.1016/j.physletb.2019.03.056;%%
  %19 citations counted in INSPIRE as of 25 May 2021
  
  %\cite{Faustov:2019ddj}
\bibitem{Faustov:2019ddj} 
  R.~N.~Faustov and V.~O.~Galkin,
  %``Semileptonic $\Xi_c$ baryon decays in the relativistic quark model,''
  Eur.\ Phys.\ J.\ C {\bf 79}, no. 8, 695 (2019)
  doi:10.1140/epjc/s10052-019-7214-5
  [arXiv:1905.08652 [hep-ph]].
  %%CITATION = doi:10.1140/epjc/s10052-019-7214-5;%%
  %7 citations counted in INSPIRE as of 25 May 2021
  
    \bibitem{Q.A.Zhang:2021}
  Q.A.Zhang, J.Hua, F.Huang, R-B.Li, Y-Y.Li, C-D.Lu, P.Sun, W.Sun, W.Wang, Y-B.Yang ,arXiv:2103.07064 [hep-ex]
  
  %\cite{Hsiao:2020gtc}
\bibitem{Hsiao:2020gtc} 
  Y.~K.~Hsiao, L.~Yang, C.~C.~Lih and S.~Y.~Tsai,
  %``Charmed $\Omega _c$ weak decays into $\Omega $ in the light-front quark model,''
  Eur.\ Phys.\ J.\ C {\bf 80}, no. 11, 1066 (2020)
  doi:10.1140/epjc/s10052-020-08619-y
  [arXiv:2009.12752 [hep-ph]].
  %%CITATION = doi:10.1140/epjc/s10052-020-08619-y;%%
  %3 citations counted in INSPIRE as of 25 May 2021
  
  %%%%%%%%%%%%%%%definition of generation form%%%%%%%%%%%%%
  %\cite{Mott:2011cx}
\bibitem{Mott:2011cx} 
  L.~Mott and W.~Roberts,
  %``Rare dileptonic decays of $\Lambda_b$ in a quark model,''
  Int.\ J.\ Mod.\ Phys.\ A {\bf 27}, 1250016 (2012)
  doi:10.1142/S0217751X12500169
  [arXiv:1108.6129 [nucl-th]].
  %%CITATION = doi:10.1142/S0217751X12500169;%%
  %42 citations counted in INSPIRE as of 25 May 2021
  

  
%%%%%%%%%%%%%lattice 
%\cite{Meinel:2016dqj}
\bibitem{Meinel:2016dqj} 
  S.~Meinel,
  %``$\Lambda_c \to \Lambda l^+ \nu_l$ form factors and decay rates from lattice QCD with physical quark masses,''
  Phys.\ Rev.\ Lett.\  {\bf 118}, no. 8, 082001 (2017)
  doi:10.1103/PhysRevLett.118.082001
  [arXiv:1611.09696 [hep-lat]].
  %%CITATION = doi:10.1103/PhysRevLett.118.082001;%%
  %29 citations counted in INSPIRE as of 25 May 2021
  
  %%%%%%%%%%%%%%%%z-expansion%%%%%%%%%%%%%
  %\cite{Bourrely:2008za}
\bibitem{Bourrely:2008za} 
  C.~Bourrely, I.~Caprini and L.~Lellouch,
  %``Model-independent description of B ---> pi l nu decays and a determination of |V(ub)|,''
  Phys.\ Rev.\ D {\bf 79}, 013008 (2009)
  Erratum: [Phys.\ Rev.\ D {\bf 82}, 099902 (2010)]
  doi:10.1103/PhysRevD.82.099902, 10.1103/PhysRevD.79.013008
  [arXiv:0807.2722 [hep-ph]].
  %%CITATION = doi:10.1103/PhysRevD.82.099902, 10.1103/PhysRevD.79.013008;%%
  %309 citations counted in INSPIRE as of 25 May 2021


  \bibitem{Commins:book} 
  E.~D.~ Commins and P.~H.~ Bucksbaum, Weak Interactions of Leptons and Quarks (Cambridge University Press, Cambridge, England, 1983).

%\cite{Meinel:2016dqj}
\bibitem{Meinel:2016dqj} 
  S.~Meinel,
  %``$\Lambda_c \to \Lambda l^+ \nu_l$ form factors and decay rates from lattice QCD with physical quark masses,''
  Phys.\ Rev.\ Lett.\  {\bf 118}, no. 8, 082001 (2017)
  doi:10.1103/PhysRevLett.118.082001
  [arXiv:1611.09696 [hep-lat]].
  %%CITATION = doi:10.1103/PhysRevLett.118.082001;%%
  %29 citations counted in INSPIRE as of 17 May 2021
  
  
  %\cite{Hsiao:2020gtc}
\bibitem{Hsiao:2020gtc} 
  Y.~K.~Hsiao, L.~Yang, C.~C.~Lih and S.~Y.~Tsai,
  %``Charmed $\Omega _c$ weak decays into $\Omega $ in the light-front quark model,''
  Eur.\ Phys.\ J.\ C {\bf 80}, no. 11, 1066 (2020)
  doi:10.1140/epjc/s10052-020-08619-y
  [arXiv:2009.12752 [hep-ph]].
  %%CITATION = doi:10.1140/epjc/s10052-020-08619-y;%%
  %3 citations counted in INSPIRE as of 25 May 2021
  
  
  %\cite{Auvil:1966eao}
\bibitem{Auvil:1966eao} 
  P.~R.~Auvil and J.~J.~Brehm,
  %``Wave Functions for Particles of Higher Spin,''
  Phys.\ Rev.\  {\bf 145}, no. 4, 1152 (1966).
  doi:10.1103/PhysRev.145.1152
  %%CITATION = doi:10.1103/PhysRev.145.1152;%%
  %71 citations counted in INSPIRE as of 26 May 2021
  

\end{thebibliography}
\end{document}